\documentclass[reprint,twocolumn,showpacs,nofootinbib,prd,amssymb]{revtex4}
\usepackage{latexsym}
\usepackage{graphicx,epsf, epsfig, amssymb, amsmath}

\usepackage[usenames,dvipsnames]{xcolor}

\usepackage{relsize}
\usepackage{dsfont}
\usepackage{todonotes}
\usepackage{textcomp}
\usepackage{float}
\usepackage{color}
\usepackage{mathrsfs}
\usepackage{soul}

\newcommand{\aap}{Astron.\ Astrophys.}
\newcommand{\mnras}{Mon.\ Not.\ R.\ Astron.\ Soc.}
\newcommand{\physrep}{Phys.\ Rep.}

\newcommand{\apjs}{Astrophys.\ J.\ Suppl.\ Ser.}

\newcommand{\Msun}{M$_\odot$}
\newcommand{\nliiiwr}{NL3$\omega\rho$}
\newcommand{\Tg}{\widetilde{T}}
\newcommand{\diff}{\mathrm{d}}
\newcommand{\pd}[2]{\frac{\partial #1}{\partial #2}}
\DeclareMathOperator{\diver}{div}
\renewcommand{\vec}[1]{\pmb{#1}}

\newcommand{\mLand}[1]{m^*_{\text{L} #1}}
\newcommand{\mLandPow}[2]{m_{\text{L} #1}^{* #2}}
\newcommand{\mDir}[1]{m^*_{\text{D} #1}}
\newcommand{\mDirPow}[2]{m_{\text{D} #1}^{* #2}}
\newcommand{\pF}[1]{p_{\text{F} #1}}
\newcommand{\zmax}{\zeta_\mathrm{max}}
\newcommand{\lmax}{\lambda_\mathrm{max}}
\newcommand{\npLp}{$np\leftrightarrow\Lambda p$}
\newcommand{\nnLn}{$nn\leftrightarrow\Lambda n$}
\newcommand{\nLLL}{$n\Lambda\leftrightarrow\Lambda\Lambda$}
\newcommand{\nXLX}{$n\Xi^-\leftrightarrow\Lambda\Xi^-$}
\newcommand{\LnXp}{$\Lambda n\leftrightarrow\Xi^-p$}

\newcommand{\ph}[1]{#1}  
\newcommand{\mg}[1]{#1}  
\newcommand{\mgdel}[1]{}  

\begin{document}

\title[Bulk viscosity in hyperon NSs]{Bulk viscosity in neutron stars with hyperon cores}

\author{D.~D.~Ofengeim$^1$\footnote{ddofengeim@gmail.com}, M.~E.~Gusakov$^1$, P. Haensel$^2$, M. Fortin$^2$} 
\affiliation{$^1$Ioffe Institute, Polytekhnicheskaya 26, 194021 St. Petersburg, Russia}
\affiliation{$^2$N. Copernicus Astronomical Center, Polish Academy of Sciences, Bartycka 18, 00-716 Warszawa, Poland}

\begin{abstract} 
It is well-known that r-mode oscillations of rotating neutron stars may be unstable with respect to the gravitational wave emission. It is highly unlikely to observe a neutron star with the parameters within the instability window, a domain where this instability is not suppressed. But if one adopts the `minimal' (nucleonic) composition of the stellar interior, a lot of observed stars appear to be within the r-mode instability window. One of the possible solutions to this problem is to account for hyperons in the neutron star core. The presence of hyperons allows for a set of powerful (lepton-free) non-equilibrium weak processes, which increase the bulk viscosity, and thus suppress the r-mode instability. Existing calculations of the instability windows for hyperon NSs generally use reaction rates calculated for the $\Sigma^-\Lambda$ hyperonic composition via the contact $W$ boson exchange interaction. In contrast, here we employ hyperonic equations of state where the $\Lambda$ and $\Xi^-$ are the first hyperons to appear (the $\Sigma^-$'s, if they are present, appear at much larger densities), and consider the meson exchange channel, which is more effective for the lepton-free weak processes. We calculate the bulk viscosity for the non-paired $npe\mu\Lambda\Xi^-$ matter using the meson exchange weak interaction. A number of viscosity-generating non-equilibrium processes is considered (some of them for the first time in the neutron-star context). The calculated reaction rates and bulk viscosity are approximated by simple analytic formulas, easy-to-use in applications. Applying our results to calculation of the instability window, we argue that accounting for hyperons may be a viable solution to the r-mode problem.
\end{abstract}

\date{\today}



\maketitle

\section{Introduction}
\label{sec:intro}

There are two well-known types of viscosities in a fluid. The shear viscosity $\eta$ comes from the momentum diffusion between fluid layers moving with  different velocities. The bulk viscosity $\zeta$ appears due to non-equilibrium reactions in the compressing and decompressing fluid~\cite{LL-VI}. 

Both these viscosities are important in numerous studies of neutron stars (NSs) \cite{GlamGual2018}, in particular, for damping of their r-mode oscillations \cite{Haskell2015}. The Rossby (or simply r-) modes are a subclass of the inertial oscillation modes, restoring force of which is the Coriolis force in a rotating star. The r-modes appear to be unstable to the gravitational wave emission due to the Chandrasekhar-Friedman-Schutz instability~\cite{Chandra1970,FriSch1978}. It is damped by the shear and bulk viscosities at low and high temperatures, respectively. The domain in the $\nu,T$ plot ($\nu$ is the rotation frequency and $T$ is the internal temperature of the NS) where the star is unstable is called the r-mode instability window. It is highly unlikely to observe a NS with $\nu$ and $T$ within it. See the reviews \cite{AK2001,Haskell2015}.

However, one meets a paradox~\cite{Haskell2015}: a lot of observed NSs in low-mass X-ray binaries (LMXBs) have their $\nu$ and $T$ in the unstable domain for NSs with the nucleonic ($npe\mu$) core composition. Namely, their typical temperatures are too hot to damp the instability by $\eta$ and too low to do it via $\zeta$. A lot of possible solutions to this paradox were proposed, mainly to introduce an additional damping mechanism. Some of them are reviewed in \cite{Haskell2015}. Here we focus on the option to modify the bulk viscosity $\zeta$ by the presence of hyperons in the NS core.

In a nucleonic core $\zeta$ is mainly provided by the modified Urca process, 
e.g. $n + n \to n + p + e + \tilde{\nu}_e$ and the inverse. In the most massive nucleonic NSs the direct Urca, $n \to p + e + \tilde{\nu}_e$ and the inverse, can operate. These non-equilibrium processes have the rates $\propto T^6\Delta\mu$ and $\propto T^4\Delta\mu$, respectively ($\Delta\mu$ is the chemical equilibrium distortions due to the fluid motions) \cite{Yak2001,HLY2000,HLY2001}. This means that at low temperatures these rates are strongly suppressed by a factor of $\sim (kT/\mu)^{4-6}$ ($\mu$ is a typical baryon chemical potential). The bulk viscosity due to these processes can damp the r-mode instability only at $T \sim 10^9 - 10^{10}\,$K, while NSs in LMXBs typically have $T \sim (0.3-1)\times 10^8\,$K. The suppression of the reaction rates due to nucleon pairing even worsens the problem \cite{HLY2000,HLY2001}. 

However, there are numerous models of the NS core equation of state (EoS) predicting the presence of hyperons (baryons with at least one strange quark) in deep layers of the core \cite{HPY2007,Vidana2015}. The most-widely used ones are the relativistic mean field (RMF) models due to their relative simplicity \cite{Glend2000}. The presence of hyperons dramatically changes the bulk viscosity. At low temperatures the main contribution to $\zeta$ comes from weak non-leptonic processes, e.g., $\Sigma^- + p \leftrightarrow n +n$ or $\Lambda + p \leftrightarrow n + p$. At $T < 10^9\,$K their typical rate $\propto T^2\Delta\mu$ is much larger than the Urca process rates. There were numerous calculations of the reaction rates of these processes and the corresponding bulk viscosity \cite{LO2002,HLY2002,vDD2004,NO2006,GK2008} in both normal and paired matter. Existing calculations of the r-mode instability windows for hyperonic NSs \cite{LO2002,ReisBon2003,NO2006} yield that the hyperonic enhancement of $\zeta$ is generally not enough to solve the r-mode paradox (except, maybe, for the most massive stars $\sim 2\,$\Msun, central regions of which may be free of baryon pairing). In the recent reviews \cite{Haskell2015,Vidana2015} it is argued that the hyperonic bulk viscosity is unable to close the instability window for the observed NSs. However, previous calculations of the instability window for hyperonic NSs should be revisited. First, they used the $\Sigma^-\Lambda$ hyperonic composition of the NS core. Various modern EoS models \cite{GHK2014,RadSedWeb2018,TolosCool2018}, in particular those, calibrated to the up-to-date hypernuclear data \cite{Fortin+2017,Prov+2018} predict that $\Lambda$ and $\Xi^-$ are likely the first hyperons that appear with growing density ($\Sigma^-$-hyperons either appear at higher densities or do not appear at all in NSs). Second, calculations of \cite{LO2002,ReisBon2003,NO2006} employed reaction rates for non-leptonic weak processes derived using the contact exchange by the $W$ boson of two baryon currents. Still, it is well-known (see, e.g., Ref.~\cite{GalReview2016}), that the most effective channel for a weak inelastic collision between a hyperon and another baryon is the meson (e.g $\pi$-meson) exchange. However,  this channel was analyzed only once in Ref.~\cite{vDD2004} to calculate $\zeta$ in the NS hyperonic core. To the best of our knowledge, the results of Ref.~\cite{vDD2004} have never been used to compute the r-mode instability window.

In the present work we revisit the bulk viscosity in a non-superfluid hyperonic NS core. We consider RMF EoS models (Sec.~\ref{sec:EoSs}), for which the  $\Lambda$ and $\Xi^-$ hyperons appear  first ($\Sigma^-$ hyperons are also present in some of our EoSs, but we focus on the $\Lambda\Xi^-$ composition for simplicity). We derive relations between $\zeta$ and the rates of the weak non-leptonic processes for an arbitrary EoS (Sec.~\ref{sec:zeta-lambda}). Then, adopting the one meson exchange weak interaction model, we calculate the rates for all weak non-leptonic processes operating in the $npe\mu\Lambda\Xi^-$ matter and responsible for the bulk viscosity (Sec.~\ref{sec:lambdas}). Simple analytic approximations are proposed for $\zeta$ and the reaction rates. We continue by applying our results to calculate the r-mode instability windows for hyperonic NSs (Sec.~\ref{sec:windows}). Our results indicate that the hyperonic solution to the r-mode paradox is likely more viable than it was thought before. 
Conclusions and some discussion are given in Sec.~\ref{sec:conc}.

\section{Modern equations of state}
\label{sec:EoSs}

\begin{table}
\begin{center}
\caption{\label{tab:astro} Parameters of key-point NS models for the used EoS models: the central baryon density, $n_b$, and energy density, $\rho$, mass $M$ and radius $R$.}
\renewcommand{\arraystretch}{1.4}
\setlength{\tabcolsep}{0.15cm}
\begin{tabular}{l|lcccc}
\hline\hline
          &                   &  $n_b$        &  $\rho$         &  $M$      &  $R$   \\ 
          &                   &  [fm$^{-3}$]  &  [$10^{14}\,$g cm$^{-3}$]  &  [\Msun]  &  [km]  \\ 
\hline 
GM1A      &  typical NS       &  0.332        &  5.92           &  1.40     &  13.72  \\
          &  $\Lambda$ onset  &  0.348        &  6.25           &  1.48     &  13.71  \\
          &  $\Xi^-$ onset    &  0.408        &  7.49           &  1.67     &  13.64  \\
          &  max mass         &  0.926        &  20.10          &  1.992    &  11.94  \\
          &  $\Xi^0$ onset    &  0.988        &  21.85          &  ---      &  ---    \\
\hline
TM1C      &  typical NS       &  0.315        &  5.63           &  1.40     &  14.31  \\
          &  $\Lambda$ onset  &  0.347        &  6.28           &  1.55     &  14.23  \\
          &  $\Xi^-$ onset    &  0.463        &  8.76           &  1.85     &  13.87  \\
          &  max mass         &  0.852        &  18.42          &  2.054    &  12.48  \\
          &  $\Xi^0$ onset    &  0.936        &  20.76          &  ---      &  ---    \\
\hline 
\nliiiwr  &  typical NS       &  0.293        &  5.16           &  1.40     &  13.73  \\
          &  $\Lambda$ onset  &  0.352        &  6.39           &  1.95     &  14.03  \\
          &  $\Xi^-$ onset    &  0.474        &  9.29           &  2.50     &  13.86  \\
          &  $\Sigma^-$ onset &  0.500        &  9.97           &  2.56     &  13.77  \\
          &  max mass         &  0.699        &  16.04          &  2.707    &  12.94  \\
\hline 
FSU2H     &  $\Lambda$ onset  &  0.328        &  5.82           &  1.38     &  13.30  \\
          &  typical NS       &  0.331        &  5.87           &  1.40     &  13.31  \\
          &  $\Xi^-$ onset    &  0.421        &  7.73           &  1.69     &  13.35  \\
          &  $\Sigma^-$ onset &  0.592        &  11.52          &  1.91     &  12.95  \\
          &  max mass         &  0.901        &  19.32          &  1.993    &  11.98  \\
\hline\hline
\end{tabular}
\end{center}
\end{table}

\begin{figure}
\begin{minipage}{0.99\columnwidth}
\includegraphics[width=\textwidth]{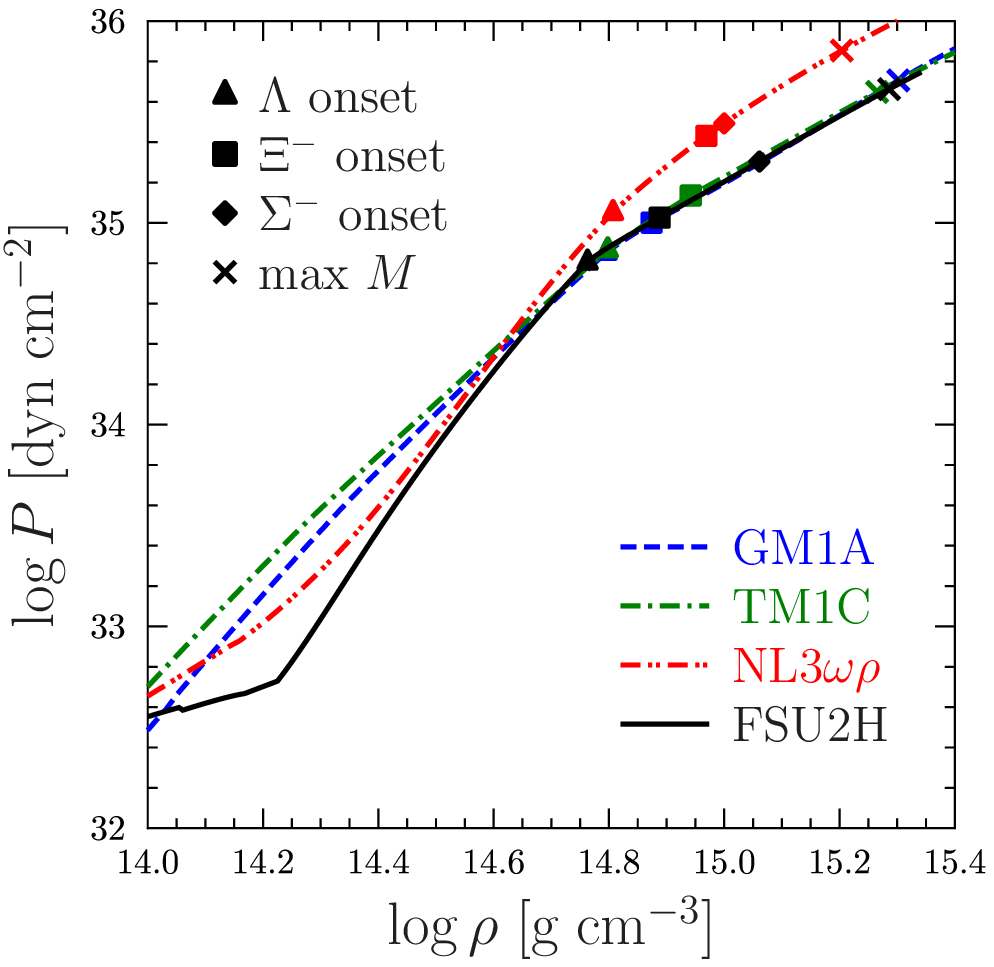}
\caption{\label{fig:P-rho} Pressure versus density for the chosen EoS models.}
\end{minipage}
\\
\vspace{0.5cm}
\begin{minipage}{0.99\columnwidth}
\includegraphics[width=\textwidth]{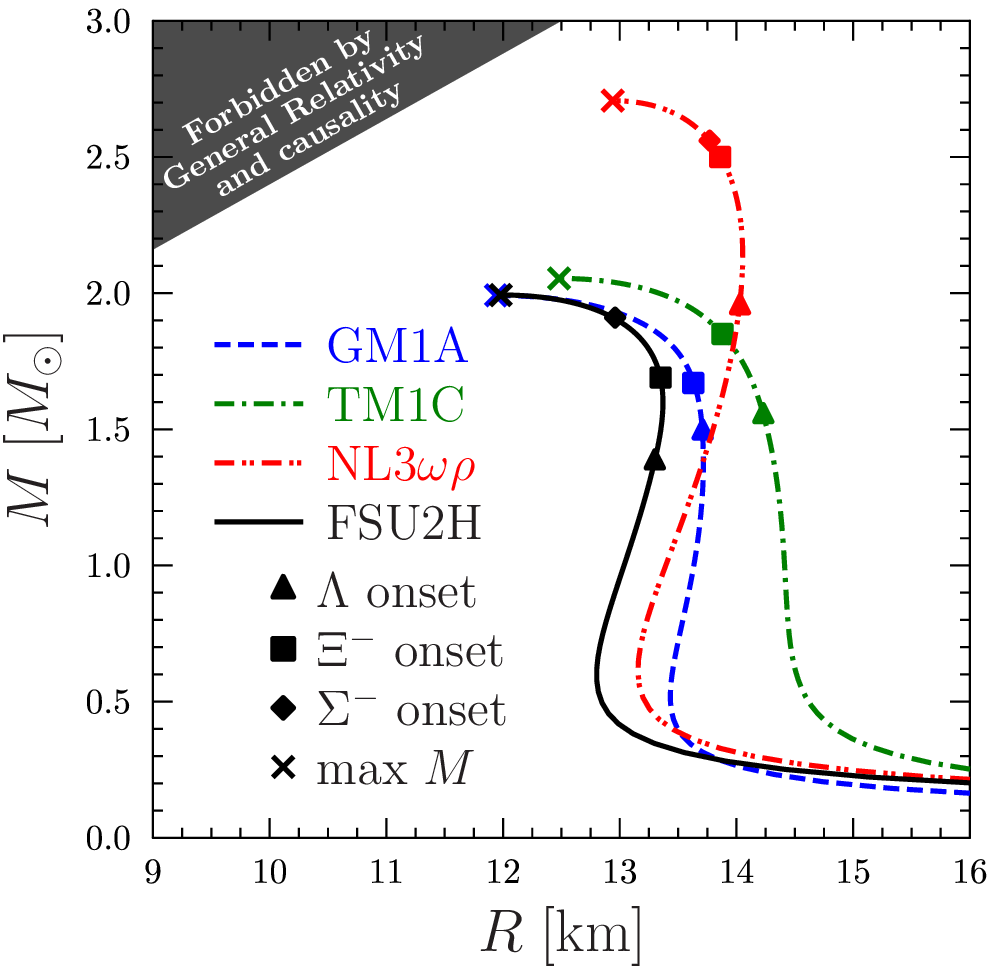}
\caption{\label{fig:M-R} Mass --- radius relations for the chosen EoS models.}
\end{minipage}
\end{figure}

Four RMF models for the core EoS are employed in this work: 
GM1A and TM1C from Ref.\ \cite{GHK2014}, \nliiiwr\ from Ref.\ \cite{Horowitz01}, and FSU2H from Ref.\ \cite{Prov+2018}. The {two last} EoSs are calibrated to the up-to-date (hyper)nuclear data {following the approach presented in Ref.~\cite{Fortin+2017}}, the former two are not. {For the FSU2H in particular we use {a} $\Sigma^-$ potential in the symmetric nuclear matter of $40$~MeV {so that  $\Sigma^-$ appear at large enough densities and masses: $M>1.9$~\Msun\ (see also the discussion in Ref.~\cite{Prov+2018})}.} {In each case, the crust EoS is calculated consistently to the core one, similarly as it was done in~\cite{Prov+2018,Fortin16}}.

The main astrophysical parameters for {the four} models are listed in Table~\ref{tab:astro}. {Fig.~\ref{fig:P-rho} shows the pressure  $P$ as a function of the density and Fig.~\ref{fig:M-R} the associated relations between the mass $M$ and the radius $R$ of NSs as obtained when solving the Tolman-Oppenheimer-Volkov equations} (e.g. \cite{Lind1992}) for these EoSs. One can see that for {the models considered here} $\Lambda$ appears first, $\Xi^-$ comes after, and then other hyperon species emerge at rather high densities and NS masses. {This} allows us to diminish the number of reactions responsible for the bulk viscosity we have to consider. In particular, within this EoS set we can limit ourselves to the properties of $npe\mu\Lambda\Xi^-$ composition up to $M\leqslant 1.9$~\Msun. 

{All models we consider are consistent with the existence of the most massive NSs with a precisely measured mass: PSR J$1614-2230$ \cite{Demorest2010,Arzoumanian18} and PSR J$0348+0432$ \cite{Antoniadis2013} with NL3$\omega\rho$ giving the largest maximum mass of all models: $\sim 2.7$~\Msun\ compared to $\sim 2$~\Msun\ for the three other paramterizations.  However only NL3$\omega\rho$ and FSU2H have values of the symmetry energy and its slope consistent with modern experimental constraints (see the discussion in e.g. \cite{Fortin16,Oertel17}). Of all models, FSU2H gives the lowest radii $R\sim 13$~km of NSs with the canonical mass $1.4$~\Msun. Note  that for the hyperonic FSU2H  EoS hyperons {are present} in NSs with a mass larger than $1.38$~\Msun.}

\begin{figure*}
\begin{center}
\includegraphics[width=0.95\textwidth]{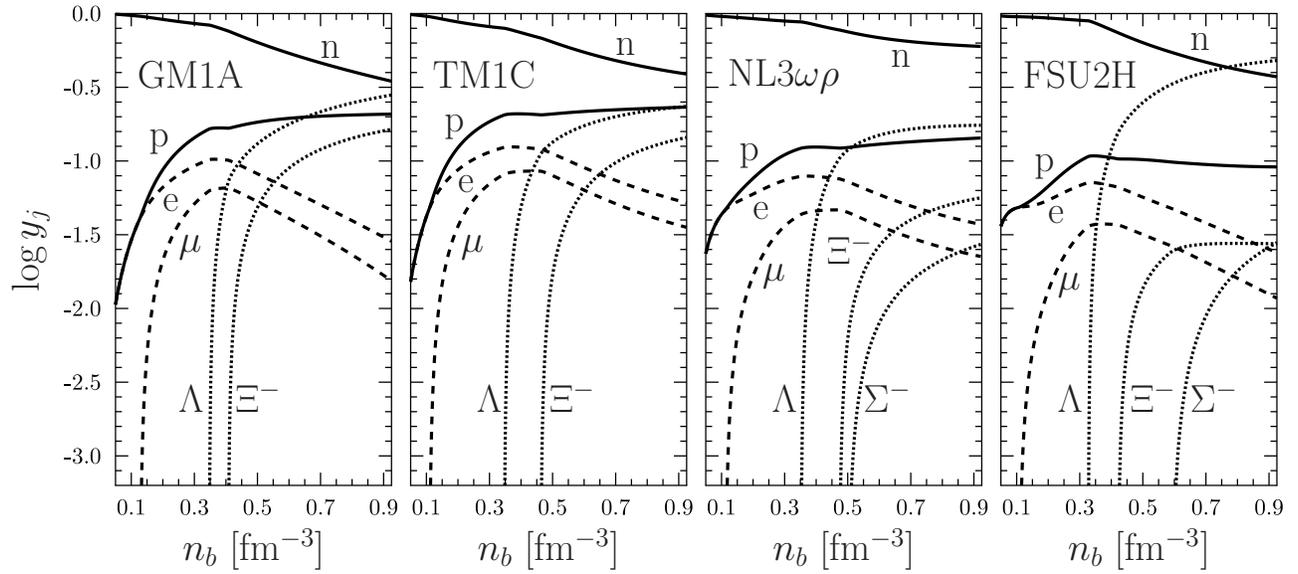}
\caption{\label{fig:yi} Particle fractions $y_j = n_j/n_b$ for species $j = n,p,e,\mu,\Lambda,\Xi^-,\Sigma^-$, which emerge in NSs of EoS models we consider.}
\end{center}
\end{figure*}

Figure~\ref{fig:yi} shows {that the four models} have significantly different composition, and we thus {expect them to give} different properties for {the} bulk viscosity.

\begin{figure*}
\begin{center}
\includegraphics[width=0.95\textwidth]{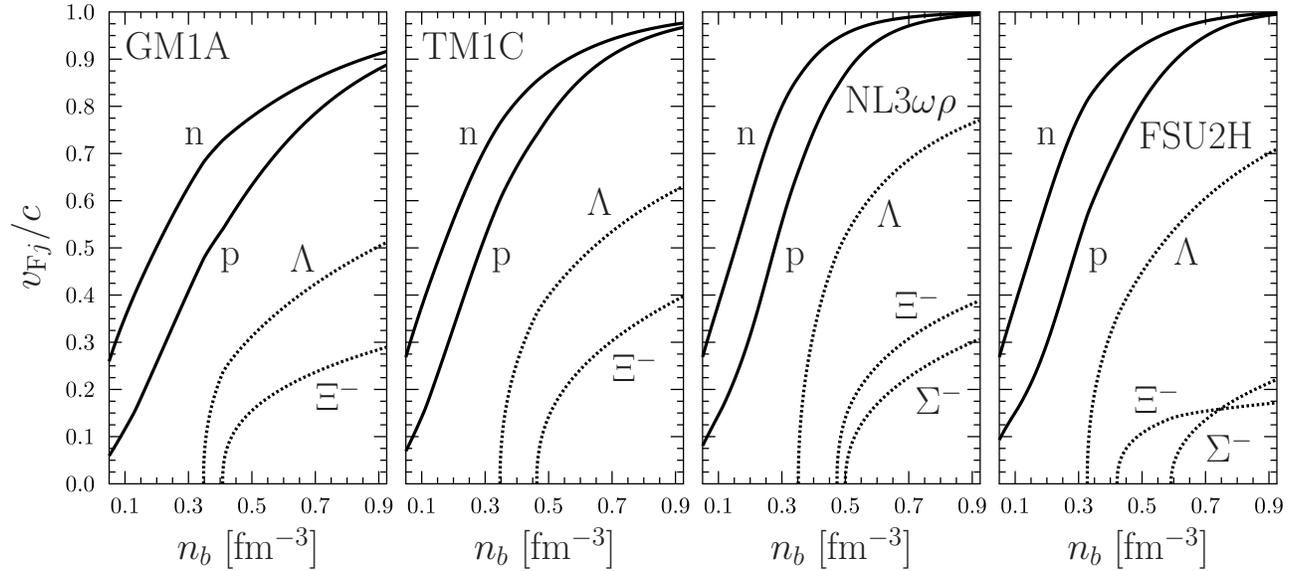}
\caption{\label{fig:relativity} Baryon Fermi velocities for baryon species $j = n,p,\Lambda,\Xi^-,\Sigma^-$, for the EoS models used.}
\end{center}
\end{figure*}

With the method presented in Ref.~\cite{GHK2014} we have calculated {the} Landau effective masses $\mLand{j}$ and Landau parameters $F_0^{jk}$ and $F_1^{jk}$ ($j$ and $k$ for all baryon species presented for a given EoS). {The quantities} $\mLand{j}$ and $F_0^{jk}$ are necessary for bulk viscosity calculations. We would like to stress that baryon Fermi velocities $v_{\text{F}j} = \pF{j}/\mLand{j}$ are close to the unity {(i.e. to the speed of light)} in a wide range of densities for all EoSs considered, see Figure~\ref{fig:relativity} for details. In other {words}, baryons (particularly nucleons) are essentially relativistic even at densities {typical} of a moderately heavy NS, $M \sim 1.5-1.9$~\Msun. Thus one has to work in the {relativistic} framework like, e.g., in Refs.\ \cite{LO2002,vDD2004,NO2006}, rather than in the nonrelativistic one (as, e.g., in Ref.\ \cite{HLY2002}),
while calculating reaction rates for the bulk viscosity.

\section{Bulk viscosity in a non-superfluid matter and reaction rates}
\label{sec:zeta-lambda}

{Bulk viscosity is generated due to non-equilibrium reactions.}
In {the case} of the nucleon $npe\mu$ matter the main reactions are the Urca processes \cite{HLY2000,HLY2001}. When {the} hyperons appear, the non-leptonic weak processes become the main source for the bulk viscosity
(see, e.g., \cite{LO2002,HLY2002}), since they are much more intensive {at typical NS temperatures}. 
There are a lot of such processes. If $\Lambda$  is the only hyperon species in the matter, the reactions are
\begin{subequations}
\label{eq:weakReact}
\begin{align}
n + p &\leftrightarrow \Lambda + p,  \label{eq:Lpnp} \\
n + n &\leftrightarrow \Lambda + n,  \label{eq:Lnnn} \\
n + \Lambda &\leftrightarrow \Lambda + \Lambda.  \label{eq:LLnL}
\end{align}
When $\Xi^-$-hyperons appear, \mg{we have two more reactions}
\begin{align}
n + \Xi^- &\leftrightarrow \Lambda + \Xi^-,  \label{eq:LXnX} \\
\Lambda + n &\leftrightarrow \Xi^- + p.  \label{eq:XpLn}
\end{align}
\end{subequations}
{The} appearance of any {additional}  hyperon species {increases} {the} number of {the relevant} processes significantly. 
Notice also that we consider only those reactions 
which change the strangeness by unity, {$\vert\Delta S\vert = 1$}. 

Non-equilibrium rates of these processes, $\Delta\Gamma_\alpha$, $\alpha = (a)$, $(b)$, $(c)$, $(d)$, $(e)$, depend on {the} chemical equilibrium perturbations $\Delta\mu_\alpha$, where, e.g., $\Delta\mu_{(a)} = \mu_n - \mu_\Lambda$, $\Delta\mu_{(e)} = \mu_\Lambda + \mu_n - \mu_{\Xi^-} - \mu_p$, etc. In the subthermal regime, $\Delta\mu_\alpha \ll kT$ ($k$ is the Boltzmann constant), the reaction rates can be written as
\begin{equation}
\label{eq:Gamma-lambda}
\Delta\Gamma_\alpha = \lambda_\alpha \Delta\mu_\alpha.
\end{equation} 
{In what followsthe quantities $\lambda_\alpha$ and $\Delta\Gamma_\alpha$ will be both referred to as ``the reaction rates''}

There are also strong hyperon reactions in the NS core. 
In the absence of pairing they are $\sim 14-16$ orders {of magnitude} 
faster than the weak non-leptonic ones. 
For NS oscillations of interest, 
with frequency $\sim 10^2 - 10^4$~Hz, 
the core matter can be considered as equilibrated with respect to them. 
{In spite of that, strong processes are also important for the bulk viscosity calculation (see below).}

There {are} no strong {hyperon}  reactions in {the} $npe\mu\Lambda$ matter. If we add $\Xi^-$, the only strong process is
\begin{subequations}
\label{eq:strongReact}
\begin{equation}
\label{eq:XpLL}
\Xi^- + p \leftrightarrow \Lambda + \Lambda.
\end{equation}
If we add $\Sigma^-$, the \ph{strong} process
\begin{equation}
\label{eq:SpLn}
\Sigma^- + p \leftrightarrow \Lambda + n
\end{equation}
{becomes available}.  Adding $\Xi^0$ we switch on the process
\begin{equation}
\label{eq:XnLL}
\Xi^0 + n \leftrightarrow \Lambda + \Lambda. 
\end{equation}
\end{subequations}
Linear combinations of these reactions are also possible. {The complete set of reactions for the full baryon octet can be found in appendix~C of \cite{GHK2014}}. 

We follow Ref.~\cite{GK2008} in {describing the recipe to derive the bulk viscosity in a form convenient for studying dissipation during NS oscillations}.

\noindent (i) Let us consider a small harmonic perturbation of the fluid with the velocity $\vec{u}$.
{It is assumed that the perturbation depends on time $t$ as 
$\propto \exp(i\omega t)$, where $\omega$ is the frequency of the perturbation.}
{The unperturbed background is taken to be in full hydrostatic and thermodynamic equilibrium.}

\noindent (ii) The fluid motion 
causes small departures $\delta n_j \propto \exp(i\omega t)$ from the equilibrium values 
{of baryon number densities,} $n_j$. 
Perturbations of chemical potentials and pressure can then be presented as
\begin{equation}
\label{eq:dP+dmu}
\delta \mu_j = \sum_k \pd{\mu_j}{n_k} \delta n_k, 
\quad 
\delta P = \sum_j n_j \delta\mu_j,
\end{equation}
where $\partial\mu_j/\partial n_k$ should be calculated near equilibrium. 
These derivatives are related to the Landau effective masses 
and Landau parameters $F_0^{jk}$ (see, e.g, equation D1 in Ref.\ \cite{GHK2014}).

\noindent (iii) The bulk viscosity $\zeta$ is defined as \cite{GK2008}
\begin{equation}
\label{eq:zetaDef}
\delta P - \delta P_\text{eq} = - \zeta \diver \vec{u}.
\end{equation}
Here $\delta P_\text{eq}$ is the pressure perturbation 
derived \ph{assuming that weak processes (\ref{eq:weakReact}) are prohibited}.\footnote{
{See \cite{LO2002} for an alternative approach to the definition of $\zeta$. 
The resulting expression for the coefficient ${\rm Re}(\zeta)$, 
which is responsible for dissipation, is the same in both approaches (as it should be).}
} {Notice that since} we use complex exponents, one {has} to calculate $\mathrm{Re}\zeta$ {when} considering dissipation.

\noindent (iv) The relation between the reaction rates and $\diver \vec{u}$ is provided by the continuity equations
\begin{equation}
\label{eq:cont-Start}
\pd{n_j}{t} + \diver n_j \vec{u} = \Delta\Gamma_j,
\end{equation}
where $\Delta\Gamma_j$ is the total {number of particles of the $j$ species produced in unit volume per unit time (reaction rate)} due to both weak and strong\footnote{
While chemical disturbance with respect to strong reactions is negligible, 
{rates of} these reactions {are} comparable to the rates of weak reactions (\ref{eq:weakReact}). 
See \cite{Jones2001,GK2008} for more details.
} reactions. These equations should be linearized with respect to $\delta n_j$ and $\vec{u}$. To calculate $\zeta$, one can neglect spatial variations of unperturbed $n_j$ (the result is applicable to both uniform and non-uniform matter, e.g. \cite{GYG2005}).

Density variations $\delta n_j$ are {linearly} dependent, because they are related by the electric neutrality condition
\begin{equation}
\label{eq:chargeless}
\sum_j e_j \delta n_j = 0
\end{equation}
($e_j$ is the electric charge of the particle species $j$)
and equilibrium conditions with respect to strong reactions 
[e.g., the reactions in Eqs~(\ref{eq:strongReact})]:
\begin{subequations}
\label{eq:strongEquil}
\begin{align}
\delta\mu_{\Xi^-} + \delta\mu_p &= 2\delta\mu_\Lambda,  \label{eq:strongEquil-XpLL} \\
\delta\mu_{\Sigma^-} + \delta\mu_p &= \delta\mu_\Lambda + \delta\mu_n,  \label{eq:strongEquil-SpLn} \\
\delta\mu_{\Xi^0} + \delta\mu_n &= 2\delta\mu_\Lambda,  \label{eq:strongEquil-XnLL}
\end{align}
\end{subequations}
etc., supplemented with Eq.~(\ref{eq:dP+dmu}) for $\delta\mu_j$. 
Therefore, for any number of particle species, only four of density perturbations $\delta n_j$ are independent. 

Another important consequence of Eqs.~(\ref{eq:strongEquil}) is that for all non-leptonic weak processes we have
\begin{equation}
\label{eq:unifiedDmu}
\Delta\mu_\alpha = \Delta\mu_{(a)} = \delta\mu_n - \delta\mu_\Lambda = \Delta\mu.
\end{equation}
This is, \mg{in particular}, true for reactions that are listed in Eqs.~(\ref{eq:weakReact}).

The most convenient choice of four independent thermodynamic parameters is: the baryon number density $n_b$ ({conserved} in all reactions), the electron and muon fractions $y_{e,\mu} = n_{e,\mu}/n_b$ ({conserved since we restrict ourselves to non-leptonic  reactions}), and the strangeness fraction $y_s = \sum_j S_j n_j/n_b$, {where} $S_j$ is the strangeness of the species $j$. {Only weak processes contribute to the strangeness production} since {it is conserved} in strong reactions. As we consider weak non-leptonic reactions with $\Delta S = 1$ only, the total strangeness production {rate} $\Delta\Gamma_S$ is just the sum of all {partial} rates $\Delta\Gamma_\alpha$. Employing Eq.~(\ref{eq:unifiedDmu}) and bearing in mind that $S_j<0$, we have
\begin{equation}
\label{eq:lambTotDef}
\Delta\Gamma_S = - \lambda \Delta\mu, \quad \lambda = \sum_\alpha \lambda_\alpha,
\end{equation}
where $\lambda$ is the total reaction rate of all non-leptonic weak processes. 

The continuity Eqs.~(\ref{eq:cont-Start}) {lead} to 
\begin{subequations}
\label{eq:cont-Fin}
\begin{align}
\delta n_b &= \frac{i}{\omega} n_b \diver \vec{u},  \label{eq:cont-Fin:nb} \\
\delta y_e &= \delta y_\mu = 0,  \label{eq:cont-Fin:ylep} \\
\delta y_s &= -\frac{i\Delta\Gamma_S}{\omega n_b} = \frac{i\lambda}{\omega n_b}\Delta\mu.  \label{eq:cont-Fin:ys}
\end{align}
\end{subequations}
Considering all thermodynamic quantities as functions of $n_b$ and $y_{e,\mu,s}$ and accounting for Eq.~(\ref{eq:cont-Fin:ylep}), we get
\begin{subequations}
\label{eq:dPDmu-dnbdy}
\begin{align}
\delta P &= \pd{P}{n_b}\delta n_b + \pd{P}{y_s}\delta y_s,  \label{eq:dP-dnbdy} \\
\Delta\mu &= \pd{\Delta\mu}{n_b}\delta n_b + \pd{\Delta\mu}{y_s}\delta y_s  \label{eq:Dmu-dnbdy}
\end{align}
\end{subequations}
with $\partial \Delta\mu/\partial X = \partial \mu_n/\partial X - \partial \mu_\Lambda/\partial X$ {stemming from Eq.~(\ref{eq:unifiedDmu})}. Near-equilibrium derivatives with respect to $n_b$ and $y_s$ can be derived from Eqs.~(\ref{eq:dP+dmu}), (\ref{eq:chargeless}), and (\ref{eq:strongEquil}). The quantity $\delta P_\text{eq}$ should be calculated with Eq.~(\ref{eq:dP-dnbdy}) {assuming that} all reactions are switched off, i.e. $\delta y_s = 0$ as well as {$\delta y_e=0$ and $\delta y_\mu=0$}. 

Combining Eqs.~(\ref{eq:zetaDef}), (\ref{eq:cont-Fin}), and (\ref{eq:dPDmu-dnbdy}) we have {(cf. the formulas~(22) in~\cite{GK2008} and~(17) in~\cite{HLY2002})}
\begin{equation}
\label{eq:zetaFin}
\mathrm{Re} \zeta = \zmax \frac{2 \lambda/\lmax}{1 + (\lambda/\lmax)^2},
\end{equation}
where
\begin{subequations}
\label{eq:zlmax}
\begin{align}
\zmax &= \frac{n_b}{2\omega} \pd{P}{y_s} \pd{\Delta\mu}{n_b} \left( \pd{\Delta\mu}{y_s} \right)^{-1},  \label{eq:zlmax:z} \\
\lmax &= n_b \omega \left( \pd{\Delta\mu}{y_s} \right)^{-1}.  \label{eq:zlmax:l}
\end{align}
\end{subequations}
Eq.~(\ref{eq:zetaFin}) shows a well-known feature of the hyperon bulk viscosity \cite{LO2002,HLY2002,vDD2004,NO2006,Haskell2015}: it has a maximum with respect to the rate of non-equilibrium processes $\lambda$. Consequently, it has a maximum with respect to temperature since {$\lambda$} grows with it. {Apart from} $\lambda$ the bulk viscosity depends on two {parameters i.e., $\zmax$ which is the maximum possible bulk viscosity, and $\lmax$ which is the optimal total reaction rate for a given oscillation frequency $\omega$}. They are  {determined}  by {the} thermodynamic properties of {the} EoS only, {and} not by reactions operating in the matter.

\begin{figure}
\includegraphics[width=\columnwidth]{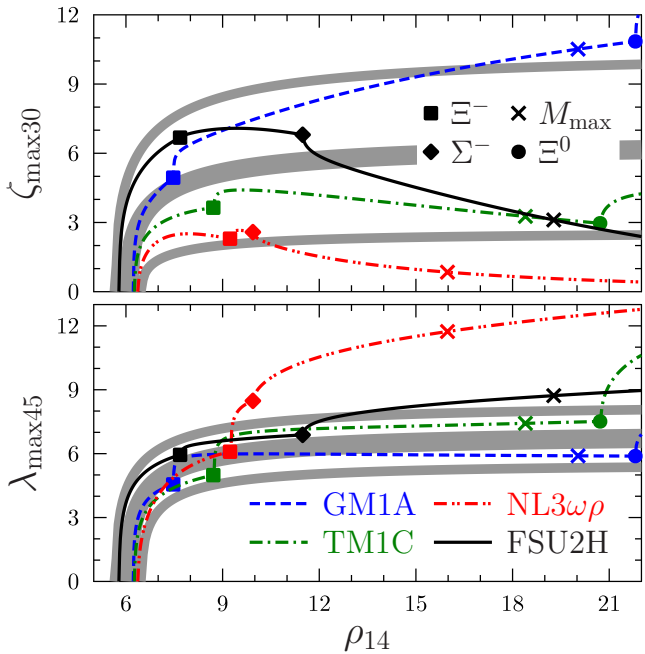}
\caption{\label{fig:zlmax} The maximum bulk viscosity $\zeta_\text{max30} = \zmax/(10^{30}\,\text{g}\,\text{cm}^{-1}\,\text{s}^{-1})$ and the optimum total reaction rate $\lambda_\text{max45} = \lmax/(10^{45}\,\text{erg}^{-1}\,\text{cm}^{-3}\,\text{s}^{-1})$ at $\omega=10^4\,$s$^{-1}$ as functions of density $\rho_{14} = \rho/(10^{14}\,\text{g}\,\text{cm}^{-3})$ for different EoS models.  Squares, diamonds, and circles mark the points of $\Xi^-$, $\Sigma^-$, and $\Xi^0$ onsets. Crosses show the state in the center of the maximum mass NS. The thicker grey lines are for the fit by Eq.~(\ref{eq:zlmaxFit}). 
The thinner grey lines show {60\%} ($\zmax$) and 20\% ($\lmax$) deviations from the fit [i.e. $\zmax^\text{appr}\times (1\pm 0.6)$ and $\lmax^\text{appr}\times(1\pm 0.2)$].}
\end{figure}

Figure~\ref{fig:zlmax} shows $\zmax$ and $\lmax$ as {functions} of energy density $\rho$. All the curves start from zero at the points of $\Lambda$ onset. The appearance of a new hyperon {causes} a rapid increment of the optimum rate $\lmax$, however, without discontinuity. The maximum viscosity $\zmax$ increases when each of cascade hyperons appears, and decreases when $\Sigma^-$ {appears}. But the main feature of plots in Figure~\ref{fig:zlmax} is that both $\zmax$ and $\lmax$ are strongly sensitive to the EoS model. However, at not too high densities, $\rho \lesssim 3\rho_0$, for all EoSs considered $\lmax(\rho)$ has similar behaviour and values.

When only $\Lambda$ and $\Xi^-$ hyperons are {present} in the core, the {averaged} behavior of the curves in Fig.~\ref{fig:zlmax} is roughly reproduced by formula
\begin{equation}
\label{eq:zlmaxFit}
\begin{pmatrix}
\zmax^\text{appr} \\
\lmax^\text{appr}
\end{pmatrix}
=
\begin{pmatrix}
\zeta_0/\omega_4 \\
\lambda_0 \omega_4
\end{pmatrix}
\left( \frac{x}{1+s x} \right)^t,
\quad
x = \frac{\rho-\rho_\Lambda}{\rho_0},
\end{equation}
where $\omega_4 = \omega/(10^4\,\text{s}^{-1})$ and $\rho_\Lambda$ is the density of $\Lambda$ hyperon onset (see Table~\ref{tab:astro}). The fitting parameters are $\zeta_0 = 6.5\times 10^{30}\,$g$\,$cm$^{-1}\,$s$^{-1}$, $\lambda_0 = 8.0\times 10^{45}\,$erg$^{-1}\,$cm$^{-3}\,$s$^{-1}$, $t = 0.34$, and $s = 1.0$ for $\zmax$ ({maximum} error {$\sim 60\%$}) and $s = 1.5$ for $\lmax$ ({maximum} error $\sim 20\%$) respectively. We emphasize that the power $t$ describing the behavior at $\rho \to \rho_\Lambda$ is the same for both these quantities. The thicker grey curves in Fig.~\ref{fig:zlmax} show how this fit works, and the {thinner ones visualize} {60\%} and 20\% {uncertainties} for $\zmax$ and $\lmax$, correspondingly. 
Of course, Eq.~(\ref{eq:zlmaxFit}) does not reproduce kinks at {the} $\Xi^-$ onset points
and it {does not describe behavior of the curves after appearance of $\Sigma^-$ or $\Xi^0$ hyperon}. 
However, the four EoSs we use here are significantly different, 
and we can hope that, for the $npe\mu\Lambda\Xi^-$ matter,
{any other RMF model would give $\zmax$ and $\lmax$ within the range of uncertainties predicted by our fit (\ref{eq:zlmaxFit}).}

When plotting r-mode instability windows, the averaged fit for $\lmax^\text{appr}$ appears to be rather accurate, but the fit for $\zmax^\text{appr}$, without additional corrections, fails to reproduce the r-mode instability window for some specific EoS. See the end of Sec.~\ref{sec:windows} and the caption to Fig.~\ref{fig:windowCheck} for a description of how one should use Eq.~(\ref{eq:zlmaxFit}) to solve this problem.

Now, the question is how close {the ``real''} reaction rate of weak non-leptonic reactions $\lambda$ {can} be to the optimum rate.

\section{Nonleptonic weak processes}
\label{sec:lambdas}

\subsection{General formalism}
\label{sec:lambdas-gen}

{The formalism of reaction rate calculation that we use follows} \cite{HLY2002,vDD2004}. In general, we {consider  a process in which  a pair  of baryons}\footnote{
Stricly speaking, in the dense  nucleon-hyperon matter  of NS cores we have to consider {`the baryon quasiparticles' instead of  `baryons', the latter being appropriate in  vacuum or in a few baryon systems. Hereafter by `baryon' or `particle'  we will mean  `the baryon quasiparticle'}.
} transforms into another one,
\begin{equation}
\label{eq:processGen}
1 + 2 \leftrightarrow 3 + 4,
\end{equation}
where for {baryon  strangenesses} the rule $|S_1 + S_2 - S_3 - S_4|=1$ holds. If the baryon composition is $np\Lambda\Xi^-$, then we {are left with} only the five processes listed in Eq.~(\ref{eq:weakReact}). 

An inelastic collision $1+2\to 3+4$ is described by a matrix element $\mathcal{M}_{12\to 34}$. Hereafter we assume that during its calculation the particle wavefunctions are normalized to one  particle per unit volume. Then, {setting} $\hbar=c=1$ and treating particles {as} non-polarized, the {expression for the rate} of a direct reaction $1+2 \rightarrow 3+4$ is 
\begin{multline}
\label{eq:Gdir}
\Gamma_{\to} = \int \prod_{j=1}^4 \frac{\diff^3 \vec{p}_j}{(2\pi)^3 2\mLand{j}} (2\pi)^4 \delta(p_1 + p_2 - p_3 - p_4) \times \\
 \frac{1}{s}\sum_\text{spins} \bigl| \mathcal{M}_{12\to 34} \bigr|^2 f_1 f_2 (1-f_3) (1-f_4),
\end{multline}
where $p_j = (\epsilon_j,\vec{p}_j)$ is a $j$'th quasiparticle 4-momentum, $s$ is the symmetry factor, which is equal to 2 for the reactions~(\ref{eq:Lnnn}) and (\ref{eq:LLnL}), otherwise $s=1$, and
\begin{equation}
\label{eq:fjDef}
f_j = f(z_j), \quad z_j = \frac{\epsilon_j - \mu_j}{kT}, \quad f(z) = \frac{1}{1+e^z}
\end{equation}
is the Fermi {distribution} function.

Since the  {fermions  in the NS core matter are} strongly degenerate, one can perform the phase space decomposition \cite{ShapTeuk1983} in (\ref{eq:Gdir}):
\begin{equation}
\label{eq:GdirDecomp}
\Gamma_{\to} = \frac{\prod_j \pF{j}}{4(2\pi)^8 s} (kT)^3 \mathcal{I}\left( \frac{\Delta\mu}{kT} \right) \mathcal{A}\mathcal{J},
\end{equation}
where $\Delta\mu = \mu_1 + \mu_2 - \mu_3 - \mu_4$ 
({recall that} Eq.~\ref{eq:unifiedDmu} {states} that all {$\Delta\mu_\alpha$} are equal in our problem). For the factors $\mathcal{I}$, $\mathcal{A}$, and $\mathcal{J}$ we have \cite{HLY2002}
\begin{subequations}
\label{eq:IAJ}
\begin{align}
\label{eq:IAJ-I}
\mathcal{I}(\xi) =& \int \prod_j \left[ \diff z_j f(z_j) \right] \delta\left(\sum_j z_j-\xi\right) = \frac{4\pi^2\xi + \xi^3}{6(1-e^{-\xi})}, 
\\
\nonumber
\mathcal{A} =& 
\int \prod_j\diff\Omega_j \delta(\vec{p}_1 + \vec{p}_2 - \vec{p}_3 - \vec{p}_4)
\\
\label{eq:IAJ-A}
&= \frac{2(2\pi)^3}{\prod_j \pF{j}}(q_\text{max} - q_\text{min}) \Theta(q_\text{max} - q_\text{min}), 
\\
\label{eq:IAJ-J}
\mathcal{J} =& \frac{1}{\mathcal{A}} \int \prod_j \diff\Omega_j \delta(\vec{p}_1 + \vec{p}_2 - \vec{p}_3 - \vec{p}_4) \left\langle \bigl| \mathcal{M}_{12\to 34} \bigr|^2 \right\rangle,
\end{align}
\end{subequations}
where $\langle \rangle$ means summation over the final spin states and averaging over the initial ones, $\Theta(x)$ is the Heaviside function, and
\begin{subequations}
\label{eq:qMinMax}
\begin{align}
\label{eq:qMinMax-min}
q_\text{min} =& \max \left\{ |\pF{1}-\pF{3}|, |\pF{2}-\pF{4}| \right\}, 
\\
\label{eq:qMinMax-max}
q_\text{max} =& \min \left\{ \pF{1}+\pF{3}, \pF{2}+\pF{4} \right\}, 
\end{align}
\end{subequations}
are the minimum and maximum momentum transfers.

An inverse reaction $3+4\rightarrow 1+2$ has the rate $\Gamma_{\leftarrow} = \Gamma_\to (\Delta\mu \to -\Delta\mu)$, so the total process rate is
\begin{equation}
\label{eq:DeltaGamma-IAJ}
\Delta\Gamma_{12\leftrightarrow 34} = \frac{\prod_j \pF{j}}{4(2\pi)^8 s} (kT)^3 \Delta\mathcal{I}\left( \frac{\Delta\mu}{kT} \right) \mathcal{A}\mathcal{J},
\end{equation}
where
\begin{equation}
\label{eq:DeltaIdef}
\Delta\mathcal{I}(\xi) = \mathcal{I}(\xi) - \mathcal{I}(-\xi) = \frac{2\pi^2}{3}\xi \left( 1 + \frac{\xi^2}{4\pi^2} \right).
\end{equation}
In the subthermal limit, $\Delta\mu \ll kT$, Eq.~(\ref{eq:DeltaGamma-IAJ}) takes the already mentioned form of Eq.~(\ref{eq:Gamma-lambda}).

{The next tasks consist in (i) deriving an expression for $\langle|\mathcal{M}|^2\rangle$ and then (ii) averaging it via the angular integrations, yielding 
in this way the formula for $\mathcal{J}$, Eq.~(\ref{eq:IAJ-J})}.

\subsection{Matrix element}
\label{sec:lambdas-Mfi}

\begin{figure}
\includegraphics[width=\columnwidth]{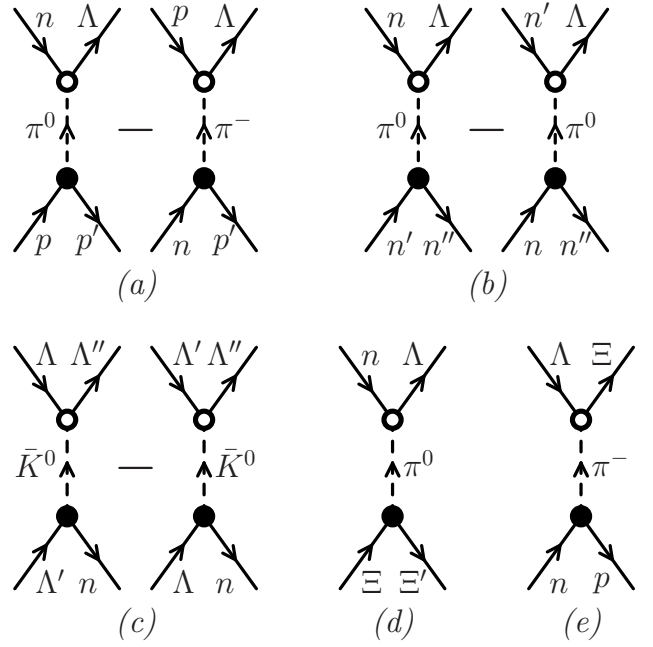}
\caption{\label{fig:diagr} The lightest meson exchange Feynman diagrams for the inelastic scatterings in Eqs.~(\ref{eq:weakReact}). Open and filled circles mark weak and strong vertices, respectively.}
\end{figure}

A non-leptonic weak reaction can go via two channels. The first one is a direct $W$-boson exchange {between}  two baryons, the weak contact interaction. The second channel is a virtual meson exchange, 
when a $W$-{boson}, emitted by one of the quarks confined in a baryon, decays into a {pair of quark and antiquark} that participate in further formation of an intermediate meson and an outgoing baryon. 

The $W$ exchange in the weak non-leptonic reactions is well-studied in context of the bulk viscosity in NS cores, e.g. \cite{LO2002,HLY2002,vDD2004,NO2006}. 

The meson-exchange channel is commonly used in studies of non-leptonic hyperon decays in laboratory, see e.g. \cite{GalReview2016} for a review. {In particular, the nucleon-induced $\Lambda$~decay and formation, \npLp\ and \nnLn, is explored in hypernuclear physics \cite{PRB1997,IM2010,BauGarPer2017} and in nucleon-nucleon scatterings \cite{Parreno+1999}. These processes are studied, e.g.,  within the one meson exchange (OME) approach, including the full pseudoscalar and vector meson octets \cite{PRB1997}, as well as with one-loop corrections \cite{PerOb+2013} and account for decay of the virtual meson into a couple of others \cite{IM2010}}. The process \nLLL\ is studied in the hyperon-induced $\Lambda$ decay in double-strange hypernuclei \cite{PRB2002,BauGarPer2015} within the OME approach. {To the best of our knowledge}, weak processes with $\Xi^-$, like \nXLX\ and \LnXp, are not studied neither experimentally nor theoretically, since the strong reactions $\Xi^-p\to\Lambda\Lambda$ and $\Xi^-n\to\Lambda\Sigma^-$ operate much more effectively. 

In general, the $W$ exchange channel for the non-leptonic hyperon decay is less effective than the meson-exchange channel. Moreover, some of the processes {have no $W$} exchange contribution due to the absence of a weak $sd$ quark current~\cite{GrKl1990}. For instance, in the set of processes (\ref{eq:weakReact}) only \npLp\ and \LnXp\ can operate with the {$W$} exchange\footnote{
This {limitation} was not so pronounced {when} the $\Sigma^-\Lambda$ hyperon composition of the core was considered \cite{HLY2002,LO2002,vDD2004,NO2006}. 
}. However, only once \cite{vDD2004} the OME channel was used for {calculating} the bulk viscosity in the NS core. Three reactions were considered in {that} work, $n n\leftrightarrow\Sigma^-p$, \npLp, and \nnLn, using both OME and $W$ exchanges. In particular, it was inferred that OME is $\sim 10$ times more intensive for \npLp. But no handy formulae were given to make  {results of  \cite{vDD2004} convenient for  applying} in further calculations involving the bulk viscosity. In {the present} work we try to {reproduce} the results of \cite{vDD2004} and adopt them to the modern hyperon {compositions} {of the NS core}.  

Considering OME, we take into account the lightest meson exchange only, the $K^0$/$\bar{K}^0$ mesons for \nLLL, and the $\pi$ mesons for the other reactions. {All} these mesons are pseudoscalar. Corresponding diagrams are shown in Fig.~\ref{fig:diagr} for each of five processes considered. {An important deficiency of our approach is that} we do not account for any other mesons, {e.g. the $\rho$ one}. Commonly, {their effect is to decrease} the reaction rate up to $3-4$ times {which} is not crucial for our purposes, see the discussion in Sec.~\ref{sec:conc}. 

\begin{table}
\begin{center}
\caption{\label{tab:constants} Phenomenological interaction constants in vertices in Fig.~\ref{fig:diagr}.}
\renewcommand{\arraystretch}{1.4}
\setlength{\tabcolsep}{0.15cm}
\begin{tabular}{lcccccc}
\hline\hline
Vertex            &   Strong $g$     &   Weak $A$   &   Weak $B$   &   Reference   \\
\hline
pp$\pi$           &   $13.3$         &   ---        &   ---        &   \cite{PRB1997}, tab. III  \\
np$\pi$           &   $13.3\sqrt{2}$ &   ---        &   ---        &   \cite{PRB1997}, tab. III  \\
nn$\pi$           &   $-13.3$        &   ---        &   ---        &   \cite{PRB1997}, tab. III  \\
$\Lambda$n$\pi$   &   ---            &   $-1.07$    &   $-7.19$    &   \cite{vDD2004}, sec. V  \\
$\Lambda$p$\pi$   &   ---            &   $1.46$     &   $9.95$     &   \cite{vDD2004}, sec. V  \\
$\Lambda$nK       &   $-14.1$        &   ---        &   ---        &   \cite{PRB1997}, tab. III  \\
$\Lambda\Lambda$K &   ---            &   $0.67$     &   $-12.72$   &   \cite{PRB2002}$^a$, tab. IV  \\
$\Xi^-\Lambda\pi$ &   ---            &   $2.04$     &   $-7.5$     &   \cite{Okun}, ch. 30.3.1  \\
$\Xi^-\Xi^-\pi$   &   $-5.4$         &   ---        &   ---        &   \cite{RSY1999}$^b$, eq. (2.14)  \\
\hline\hline
\end{tabular}
\end{center}
\begin{flushleft}
$^a$ They use the opposite sign for $\gamma^5$. \\
$^b$ Their strong $f$ couplings are related to $g$ couplings as $g = f (m_2 + m_4)/m_\pi$.
\end{flushleft}
\end{table}

There is one weak (marked by $\circ$) and one strong (marked by $\bullet$) vertex for the baryon-meson interaction in each diagram. Both weak and strong vertices are phenomenological. For the pseudoscalar meson exchange they correspond to, {respectively,} 
\begin{equation}
\label{eq:vertices}
\circ = G_\text{F} m_\pi^2 (A + B\gamma^5), \quad \bullet = g \gamma^5,
\end{equation}
where $G_\text{F} = 1.436\times 10^{-49}\,\text{erg}\,\text{cm}^3$ is the Fermi coupling constant, $m_\pi$ is the charged pion mass, and $\gamma^5 = -i \gamma^0 \gamma^1 \gamma^2 \gamma^3$. The phenomenological constants $g$, $A$, and $B$ for the vertices {in the diagrams} in Fig.~\ref{fig:diagr} are listed in Tab.~\ref{tab:constants}. Some of {these constants} are measured in laboratory, {while some are evaluated}  theoretically. 

The meson propagator $D_M(q)$, where $q$ is the 4-momentum transfer, is discussed in Sec.~\ref{sec:lambdas-gator}.

Wavefunctions of the ingoing and outgoing quasiparticles are considered within the RMF approach, {i.e.}, they {have the form of} relativistic bispinors,
\begin{equation}
\label{eq:psi}
\psi_j = C_j u_j e^{i p_j^\mu x_{j\mu}}.
\end{equation}
For strongly degenerate baryons in the NS core one can use the approximation $|\vec{p}_j|=\pF{j}$. Further, for the bispinor $u_j$ one should use $\mLand{j}$ instead of $\epsilon_j$ and the Dirac effective mass $\mDir{j}$ instead of the rest mass $m_j$. The Landau and Dirac effective masses are related by the formula \cite{Glend2000}
\begin{equation}
\label{eq:mL-mD}
\mLandPow{j}{2} = \pF{j}^2 + \mDirPow{j}{2}.
\end{equation}
Then for the normalization {constants} $C_j$ (one particle per unit volume) and the bispinor $u_j$ one {obtains}
\begin{subequations}
\label{eq:CandU}
\begin{align}
\label{eq:CandU-C}
&C_j = \frac{1}{\sqrt{2\mLand{j}}}, \\
\label{eq:CandU-uu-S}
&\bar{u}_j u_j = 2 \mDir{j}, \\
\label{eq:CandU-uu-M}
&\sum_\text{spins} u_j \bar{u}_j = \gamma^0 \mLand{j} - \vec{\gamma}\cdot\vec{p}_j + \mDir{j}.
\end{align}
\end{subequations}
{Let us notice that}  a quasiparticle dispersion relation $p_j^0 = \epsilon_j(\vec{p}_j)$ is more complex than the free particle {one}, in particular $\epsilon_j(\pF{j}) = \mu_j \neq \mLand{j}$. 

{The \npLp, \nnLn, and \nLLL\ processes involve direct and  exchange diagrams.} {However, the \nXLX\ and \LnXp\ processes do not involve exchange diagrams due to, for example, the rule $|\Delta S|=1$ which holds in each {weak} vertex}\footnote{
Strictly speaking, diagrams with permuted particles 1 and 2 would  appear if we included the next to the lightest meson.
}.
In what follows, for a process in the general form (\ref{eq:processGen}) we consider the direct and exchange diagrams {that} differ by $1\leftrightarrow 2$ permutation, with weak vertices $1,3$ and $2,3$. 

For the direct diagram one has
\begin{equation}
\label{eq:MfiDir}
\mathcal{M}_{12\to 34}^\text{dir} = G_\text{F} m_\pi^2 \bar{u}_3 (A_{13} + B_{13}\gamma^5)u_1 D_M(q) \bar{u}_4 g_{24}\gamma^5 u_2
\end{equation}
The exchange diagram corresponds to $\mathcal{M}_{12\to 34}^\text{exch} = \mathcal{M}_{12\to 34}^\text{dir}\bigr|_{1\leftrightarrow 2}$, and the total matrix element is $\mathcal{M}_{12\to 34} = \mathcal{M}_{12\to 34}^\text{dir} - \mathcal{M}_{12\to 34}^\text{exch}$. If there is no exchange diagram for the process considered, one should (artificially) set $A_{23} = B_{23} = g_{14} = 0$. 

After {averaging} over the initial and summing over the final spin states of the squared $\mathcal{M}_{12\to 34}$ we get
\begin{multline}
\label{eq:MfiSqAvSum}
\left\langle \left| \mathcal{M}_{12\to 34} \right|^2 \right\rangle = G_\text{F}^2 m_\pi^4 \left[ X(q) D_M^2(q) \right. \\
\left. + X'(q') D_M^2(q') + Y(q,q') D_M(q)D_M(q') \right],
\end{multline}
where 
\begin{equation}
\label{eq:qqP}
q = p_3 - p_1, 
\quad 
q' = p_3 - p_2,
\end{equation}
and 
\begin{subequations}
\label{eq:XYZ}
\begin{align}
\label{eq:XYZ-X}
X(q) = X(|\vec{q}|^2) = m_M^4 &X_0 + m_M^2 X_1 |\vec{q}|^2 + X_2 |\vec{q}|^4, \\
\nonumber
Y(q,q') = Y(|\vec{q}|^2 ,|\vec{q}'|^2) = m_M^4 &Y_0 + m_M^2 Y_1 |\vec{q}|^2 \\
\label{eq:XYZ-Y}
+&\ m_M^2 Y_2 |\vec{q}'|^2 + Y_3 |\vec{q}|^2 |\vec{q}'|^2, \\
\label{eq:XYZ-Z}
X'(q) = X'(|\vec{q}|^2) = m_M^4 &X'_0 + m_M^2 X'_1 |\vec{q}'|^2 + X'_2 |\vec{q}'|^4,
\end{align}
\end{subequations}
with dimensionless $X_k$, $X'_k$, and $Y_k$ being functions of $\pF{1...4}$ listed in Appendix~\ref{app:Mfi}.

The last issue {to be resolved} before we can evaluate Eq.~(\ref{eq:IAJ-J}) is to define meson propagators $D_M$.

\subsection{Meson propagators}
\label{sec:lambdas-gator}

In general, the meson propagator is
\begin{equation}
\label{eq:DM-Gen}
D_M^{-1}(\omega,\vec{q}) = \omega^2 - \vec{q}^2 - m_M^2 - \Pi_M(\omega,\vec{q}),
\end{equation}
where $\omega$ and $\vec{q}$ are the energy and momentum transferred by the virtual meson, $m_M$ is the bare (vacuum) meson mass ($m_\pi = 139\,$MeV and $m_K = 494\,$MeV)\footnote{
We do not discriminate between {masses of different members of isomultiplets}, and use values as in \cite{Glend2000}.
}, and $\Pi_M$ is the meson polarisation operator. 

Within a widely used free meson approach \cite{FM1979,Maxwell1987,vDD2004} the polarisation operator is $\Pi_M = 0$ and $\omega^2$ is omitted due to some reasons. In the almost beta-equilibrated matter of the NS core we indeed have $\omega = 0$ for neutral mesons, but {for} the charged pions in the diagrams for the processes \npLp\ (Fig.~\ref{fig:diagr}a) and \LnXp\ (Fig.~\ref{fig:diagr}e) {we} have $\omega = \mu_e \ne 0$. Thus the approach by \cite{vDD2004} to the meson propagator has to be revisited.

If we substitute $\omega = \mu_e$ into the free pion propagator, we {get into trouble as soon as} $\mu_e > m_\pi$ at $n_b \gtrsim 0.2\,$fm$^{-3}$, and the pion propagator can be positive at some {real values of momentum transfer}. This means {that the real} pions appear in the matter, but it is inconsistent with our EoS models, which (artificially) prohibit pionization. {This troubling feature} appears not only for all four EoSs {that we are using} (see Sec.~\ref{sec:EoSs}), but also for a number of other realistic nucleon EoS models like APR \cite{APR1998} and BSk21 \cite{BSK2013}. Therefore we {are forced}  to account for the polarisation operator $\Pi_{\pi^-}$ of negative pions {hoping} that at $\omega=\mu_e$ it is large enough to make $D_{\pi^-}<0$ for all densities. 

{We find it convenient} to introduce the ``effective'' virtual pion mass,
\begin{equation}
\label{eq:mPiEff}
\tilde{m}_{\pi-} = \sqrt{m_\pi^2 - \mu_e^2 + \Pi_{\pi^-}(\mu_e,\vec{q})}.
\end{equation}
Then the propagator takes a simple form
\begin{equation}
\label{eq:DpiNonrel}
D_{\pi^-}^{-1} = - \vec{q}^2 - \tilde{m}_{\pi^-}^2(\vec{q}).
\end{equation}
Notice that $\mu_e$ varies with density, so $\tilde{m}_{\pi^-}$ technically depends not only on the momentum transfer $\vec{q}$ but also on $n_b$. Obviously, $\tilde{m}_{\pi^-}$ should be strictly real when the {appearance of real pions (pionization)} is prohibited. 

In {nuclear matter characteristic of atomic nuclei} we have \cite{KKW2003} $\Pi_{\pi^-} = \Pi_S + \Delta\Pi_S + \Pi_P$, where $\Pi_S$ comes from the s-wave $n\pi$-scattering, $\Delta\Pi_S$ comes from the s-wave absorption and $\Pi_P$ is the p-wave contribution. Only $\Pi_S$ is positive, so we focus on it {in order to get} an upper estimate of $\Pi_{\pi^-}$. The {leading-order} contribution to $\Pi_S$ in the nucleon-hyperon NS core comes from the terms \cite{Kolom-Private}
\begin{equation}
\label{eq:PiS-hypCore}
\Pi_S(\omega) = \frac{\omega}{f_\pi^2}\sum_j (-I_{3j}) n_j + \frac{\sigma_N}{f_\pi^2}\left( \frac{\omega^2}{m_\pi^2} - 1 \right) n_b,
\end{equation}
where $j$ is the baryon index, $I_{3j}$ is the isospin projection of the $j^\text{th}$ baryon, $f_\pi = 92.4\,$MeV and $\sigma_N \approx 45\,$MeV. In the nucleonic matter Eq.~(\ref{eq:PiS-hypCore}) {coincides} with equation~(11) of \cite{KKW2003}. 

\begin{figure}
\includegraphics[width=\columnwidth]{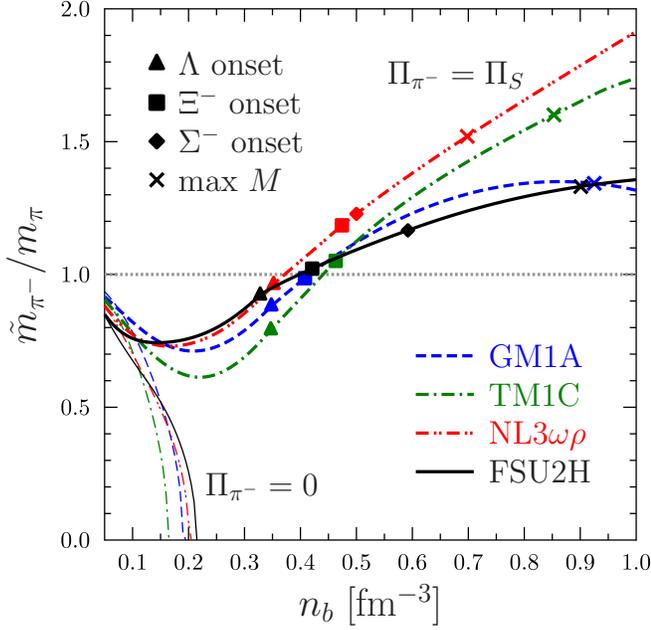}
\caption{\label{fig:mPiEff} Thick lines show the upper estimate of the `effective' pion mass for the EoS models employed. Thin lines show what happens if we do not account for the polarization operator in Eq.~(\ref{eq:mPiEff}).}
\end{figure}

Thick curves in Fig.~\ref{fig:mPiEff} show the ratio $\tilde{m}_{\pi^-}/m_\pi$ with $\Pi_{\pi^-} = \Pi_S$ for the EoS models we use in this work. Notice that in this case, according to Eq.~(\ref{eq:mPiEff}), $\tilde{m}_{\pi^-}$ technically depends on $n_b$ only. Thin curves are for $\tilde{m}_{\pi^-}$ with $\Pi_{\pi^-} = 0$. They prove what was claimed in the beginning of this section: $\mu_e$ {exceeds} the bare pion mass at $n_b \sim 0.2\,$fm$^{-3}$, so we have to account for the polarization operator to avoid a {pionization instability}. 

The s-wave part is only an upper estimate of $\Pi_{\pi^-}$, so actual values of $\tilde{m}_{\pi^-}/m_\pi$ are located below the thick lines in Fig.~\ref{fig:mPiEff}. For densities between the hyperon onset point and the maximum mass point the upper limit for $\tilde{m}_{\pi^-}$ varies in the range $(0.7...1.6)m_\pi$. Thus $m_\pi$ is a rough upper limit for $\tilde{m}_{\pi^-}$. Consequently, $1/D_{\pi^-} = -\vec{q}^2 - m_\pi^2$ is a rough lower estimate for the propagator modulus. It can be used for making a lower estimate of the reaction rates. An account for the variation of the $\tilde{m}_{\pi^-}$ upper limit mentioned above can affect a rate value not more than by a factor of order 2, which is {acceptable} for our purposes. 

Of course, accounting for other terms in $\Pi_{\pi^-}$ may dramatically {change $D_{\pi^-}$ compared to the prediction from simple expression~(\ref{eq:DpiNonrel})} with $\tilde{m}_{\pi^-} = m_\pi$. Then ``the effective pion mass'' should be {replaced} by the effective pion gap \cite{Migdal+1990}, which can be much less than $m_\pi$. Correspondingly, the pion propagator {would increase}. However, these effects are {model-dependent}, {so we prefer to use Eq.\ (\ref{eq:DpiNonrel}) with $\tilde{m}_{\pi^-} = m_\pi$ in what follows, {similarly to how it was done in}  \cite{FM1979,Maxwell1987,vDD2004}.}

What should we do with propagators of neutral mesons, {$\bar{K}^0$} and $\pi^0$? The former one is {a quite} heavy meson, and {it is harder to affect its propagator essentially}. Thus {$\bar{K}^0$} can be safely described by a free-{particle} propagator. The latter {meson}, $\pi^0$, requires more careful discussion, but one can artificially set the free-particle propagator {for} it {within} the same range of reliability as for $\pi^-$. 

All in all, for each meson propagator we use
\begin{equation}
\label{eq:DMforUse}
D_M^{-1} = -\vec{q}^2 - m_M^2.
\end{equation}
This can {lead to underestimating the reaction rates}. But this effect will be (partially) compensated by neglecting {the contribution due to  the vector mesons}, see Sec.~\ref{sec:conc} for a more detailed discussion.

\subsection{Reaction rates}
\label{sec:lambdas-rates}

Taking $\langle\left| \mathcal{M}_{12\to 34} \right|^2\rangle$ from Eq.~(\ref{eq:MfiSqAvSum}), $D_M$ from Eq.~(\ref{eq:DMforUse}), and substituting them into Eq.~(\ref{eq:IAJ-J}), we can {calculate} $\mathcal{J}$ (see Appendix~\ref{app:Iang} for details) and, consequently, {get} the reaction rate $\Delta\Gamma_{12\leftrightarrow 34}$ from Eq.~(\ref{eq:DeltaGamma-IAJ}). In the subthermal regime, $\Delta\mu\ll kT$, it can be expressed in terms of $\lambda_{12\leftrightarrow 34}$ (see Eq.~(\ref{eq:Gamma-lambda}))
\begin{equation}
\label{eq:lambda1234}
\lambda_{12\leftrightarrow 34} = \lambda_0^{12\leftrightarrow 34} \mathcal{W}_{12\leftrightarrow 34},
\end{equation}
where, restoring natural units,
\begin{subequations}
\label{eq:lambda0+Wdef}
\begin{multline}
\label{eq:lambda0}
\lambda_0^{12\leftrightarrow 34} = \frac{G_\text{F}^2 m_N^4}{6\pi^3 \hbar^{10}} \left( q_\text{max} - q_\text{min} \right) (kT)^2 \Theta_{12\leftrightarrow 34}
\\
\approx \frac{1.7\times 10^{45}}{\text{erg cm$^3$ s}} \times \frac{q_\text{max} - q_\text{min}}{\hbar \left( 3\pi^2 n_0 \right)^{1/3}} T_8^2 \Theta_{12\leftrightarrow 34},
\end{multline}
with the nucleon mass\footnote{
It is introduced here just {to} make $\mathcal{W} \lesssim 1$.
} $m_N = 939\,$MeV, $T_8 = T/(10^8\,\text{K})$, $\Theta_{12\leftrightarrow 34} = \Theta(q_\text{max} - q_\text{min})$, and
\begin{multline}
\label{eq:Wdef}
\mathcal{W}_{12\leftrightarrow 34} = \frac{1}{s}\left( \frac{m_\pi}{2m_N} \right)^4 \left( X_0 J_0 + X_1 J_1 + X_2 J_2 \right.
\\
\left. + X'_0 J'_0 + X'_1 J'_1 + X'_2 J'_2 \right.
\\
\left. + Y_0 J_3 + Y_1 J_4 + Y_2 J'_4 + Y_3 J_5, \right)
\end{multline}
\end{subequations}
is a dimensionless function of $\pF{1},\pF{2},\pF{3},\pF{4}$, with $X_k$, $X'_k$, and $Y_k$ defined in Appendix~\ref{app:Mfi}, and $J_k$ and $J'_k$ defined in Appendix~\ref{app:Iang}. Actually, $\mathcal{W}$ is related to $\mathcal{J}$ in a simple way:
\begin{equation}
\label{eq:J-W}
\mathcal{J} = 16 s G_\text{F}^2 m_N^4 \mathcal{W}.
\end{equation}
In the suprathermal regime, $\Delta\mu\gtrsim kT$, one has to use 
\begin{equation}
\label{eq:DeltaGamma1234}
\Delta\Gamma_{12\leftrightarrow 34} = \lambda_{12\leftrightarrow 34}\Delta\mu \left[ 1 + \left( \frac{\Delta\mu}{2\pi kT} \right)^2 \right].
\end{equation}

\begin{figure*}
\includegraphics[width=0.95\textwidth]{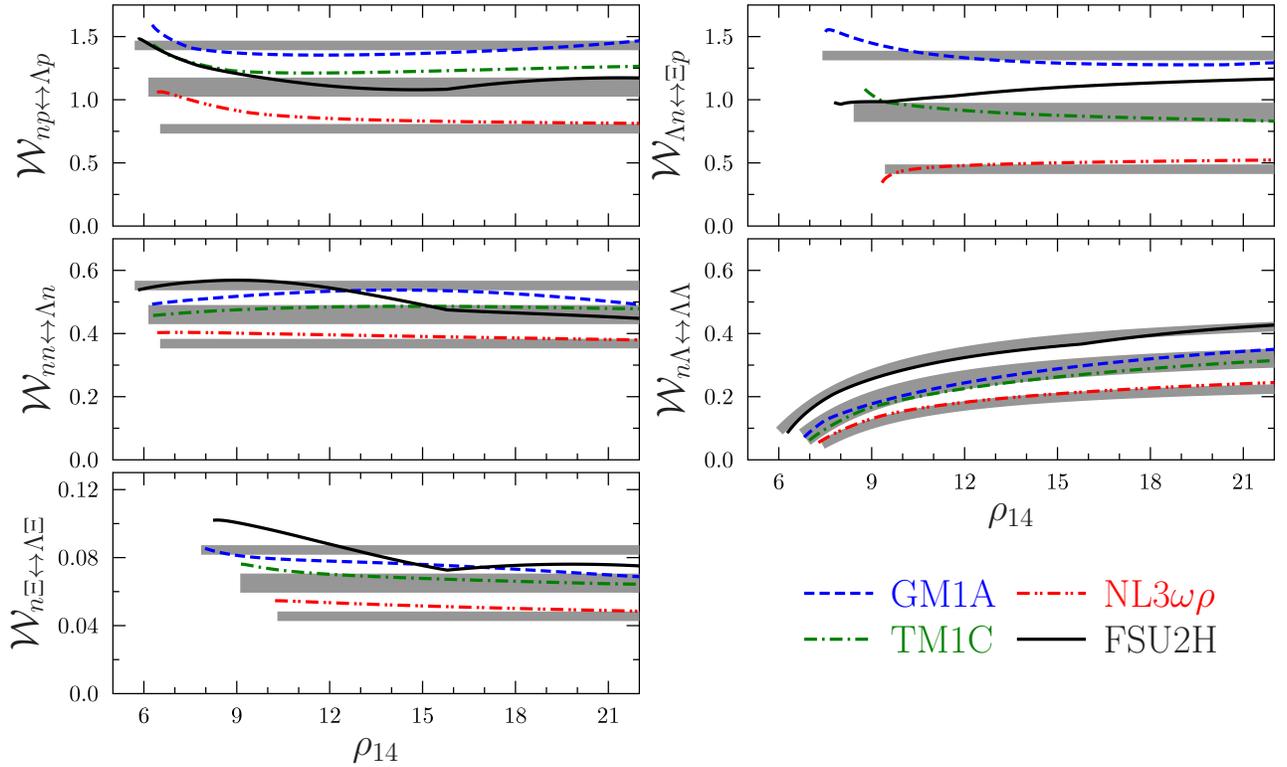}
\caption{\label{fig:W-rho} The $\mathcal{W}$ functions for the non-leptonic weak processes from Eq.~(\ref{eq:weakReact}), for the EoS models used. The thicker grey lines show $\mathcal{W}_\text{appr}$ from (\ref{eq:Wapprox}) with parameters from Table~\ref{tab:Wapprox}, the thinner ones show deviations from $\mathcal{W}_\text{appr}$ that cover most of the curves.}
\end{figure*}

\begin{table}
\begin{center}
\caption{\label{tab:Wapprox} Fitting parameters in Eq.~(\ref{eq:Wapprox}) we recommend for using in practice.}
\renewcommand{\arraystretch}{1.4}
\setlength{\tabcolsep}{0.2cm}
\begin{tabular}{lcccccc}
\hline\hline
Process   &   $W_0$  &  $a$  &  $b$  &  $p$  &  error  \\
\hline
\npLp     &   1.1    &  ---  &  ---  &  ---  &  30\%  \\
\LnXp     &   0.9    &  ---  &  ---  &  ---  &  50\%  \\
\nnLn     &   0.48   &  ---  &  ---  &  ---  &  20\%  \\
\nLLL     &   0.38   &  0.37 &  0.87 &  2    &  30\%  \\
\nXLX     &   0.068  &  ---  &  ---  &  ---  &  30\%  \\
\hline\hline
\end{tabular}
\end{center}
\end{table}

The $\mathcal{W}$ function incorporates all specific properties of the process $12 \leftrightarrow 34$ (recall that $X_k$, $Y_k$, etc. depend on weak and strong coupling constants that are different for different processes). Fig.~\ref{fig:W-rho} shows how it depends {on the 
(energy) density $\rho$} for each kind of processes in Eq.~(\ref{eq:weakReact}) for all EoSs we use. It appears to be strongly model-dependent: $\mathcal{W}$ varies up to a factor of $3$ from one EoS to another. Fortunately, it {appears} to be a slow function of $\rho$. Since the main 
{aim} of our calculations is {application} in {the} r-mode physics, 
it is enough to provide a simple (even if not too precise) approximation of the reaction rate. For \npLp, \nnLn, \LnXp, and \nXLX\ processes we can reliably treat $\mathcal{W}$ as a constant, while for \nLLL\ it is safer {to} account {that it grows} with $\rho$. The approximation {that} we recommend is
\begin{equation}
\label{eq:Wapprox}
\mathcal{W}_\text{appr} = W_0 \left( \frac{x+a}{x+b} \right)^p, \quad x = \frac{\rho - \rho_\text{start}}{\rho_0},
\end{equation}
where $\rho_\text{start}$ is the density where the process $12\leftrightarrow 34$ {switches} on, and $\rho_0 = 2.8\times 10^{14}\,\text{g}\,\text{cm}^{-3}$ is the nuclear matter saturation density. Note that $\rho_\text{start}$ may not coincide with the density of $\Lambda$ or $\Xi^-$ onset, and should be derived as a lowest density where $\Theta_{12\leftrightarrow 34}>0$. Parameters $W_0$, $a$, $b$, and $p$ represent a very rough fit of what we have in Fig.~\ref{fig:W-rho}. The latter three are required for \nLLL\ only, other processes can be described with a single constant $W_0$. In Table~\ref{tab:Wapprox} we give the parameters of this fit for each process. The thicker grey lines in Fig.~\ref{fig:W-rho} show how these fits work. The `error' column in Table~\ref{tab:Wapprox} represents `ranges of deviations', $|\mathcal{W}-\mathcal{W}_\text{appr}|/\mathcal{W}_\text{appr}$. Most of $\mathcal{W}$ curves lie within these ranges (we stress that {it is more} important to reproduce $\mathcal{W}$ behavior far from $\rho_\text{start}$ than close to it). In Fig.~\ref{fig:W-rho} the {thinner} grey lines display {boundaries of these error ranges.}
\begin{table}
\begin{center}
\caption{\label{tab:lambdaApprox} Fitting parameters in Eq.~(\ref{eq:lambdaApprox}) we recommend to use.}
\renewcommand{\arraystretch}{1.4}
\setlength{\tabcolsep}{0.3cm}
\begin{tabular}{lccccc}
\hline\hline
Process   &  $l_0$  &  $c$   &  $q$  &  error  \\
\hline
\npLp     &   1.7   &  0.06  &  0.36  &  20\%  \\
\LnXp     &   1.5   &  0.00  &  0.36  &  30\%  \\
\nnLn     &   2.9   &  0.3   &  0.4   &  20\%  \\
\nLLL     &   3.5   &  0.8   &  1.0   &  30\%  \\
\nXLX     &   1.6   &  0.5   &  1.0   &  40\%  \\
\hline\hline
\end{tabular}
\end{center}
\end{table}

\begin{figure*}
\includegraphics[width=0.95\textwidth]{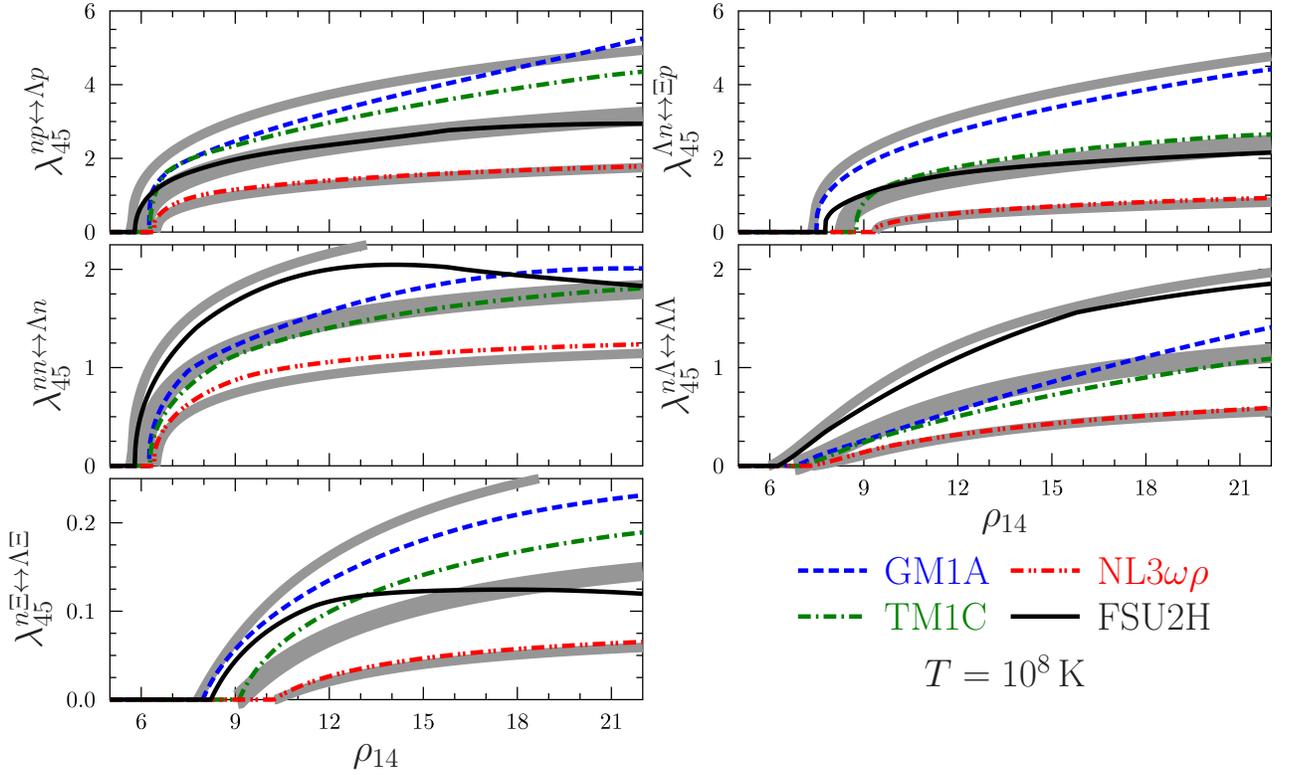}
\caption{\label{fig:lambda-rho} The reaction rates $\lambda_{45} = \lambda/(10^{45}\,\text{erg}^{-1}\,\text{cm}^{-3}\,\text{s}^{-1})$ for different EoS models at $T = 10^8\,$K. The thicker grey lines show $\lambda_\text{appr}$ from Eq.~(\ref{eq:lambdaApprox}) with best-fit parameters from Tables~\ref{tab:Wapprox} and \ref{tab:lambdaApprox}. The thinner ones show the deviations within errors from these Tables taken together, which allow to cover the whole domains occupied by $\lambda(\rho)$ curves for EoS models.}
\end{figure*}

Thus, {in order to quickly estimate}  reaction rates for an arbitrary EoS, 
one can take $\mathcal{W}$ from Eq.~(\ref{eq:Wapprox}) 
and substitute it into Eq.~(\ref{eq:lambda1234}) to obtain $\lambda$ for the process considered. The {quantity} $\lambda_0$ can be easily calculated for each process when {the number density} $n_j$ of each particle species is known. However, one may desire an approximate formula that does not require knowledge of particle fractions, e.g. to explore some phenomenological $P(\rho)$ models, supplemented with {an arbitrarily} chosen $\rho_\text{start}$. 
{For that purpose,} we provide an approximate expression for $\lambda_0$ that depends on $\rho$ and $\rho_\text{start}$ only,
\begin{equation}
\label{eq:lambdaApprox}
\lambda_{0\,\text{appr}} = l_0 \left( \frac{x}{1 + c x} \right)^q T_8^2,
\;\;\; 
\lambda_\text{appr} = \lambda_{0\,\text{appr}} \mathcal{W}_\text{appr},
\end{equation}
with the same $x$ as in Eq.~(\ref{eq:Wapprox}). Recommended values of $c$, $q$, and $l_0$ and maximum relative deviations for each process are given in Table~\ref{tab:lambdaApprox}. 

Fig.~\ref{fig:lambda-rho} shows {the density dependence $\lambda(\rho)$} for all five processes {that} we consider for EoS models from Sec.~\ref{sec:EoSs} at $T = 10^8\,$K. Grey lines show $\lambda_\text{appr}$ (thicker {lines}) and {boundaries of its uncertainty (thinner lines)} 
due to both $\mathcal{W}$ and $\lambda_0$ approximation errors. For instance, for \npLp\ the thinner lines correspond to $\lambda_{0\,\text{appr}}^\text{\npLp}\times(1\pm 0.3)\times \mathcal{W}_\text{appr}^\text{\npLp}\times(1\pm 0.2)$. The reaction rates are also model-dependent, {similarly to the $\mathcal{W}$ functions}. There is an explicit hierarchy\footnote{
We emphasize that in the superfluid matter the hierarchy is different.
} of $\lambda$ typical values. The {processes} \npLp\ and \LnXp\ {turn out} to be the most effective. The next are \nnLn\ and \nLLL. The latter one has {stronger}  $\rho$ dependence since it is more sensitive to the $\Lambda$ {fraction}. The least intensive is {the \nXLX\ process}. There are two reasons for this. First, it is most sensitive to low $\Xi^-$ density. Second, it has the lowest $B$ and $g$ coupling constants (see Tab.~\ref{tab:constants}), and it has no exchange term contribution in our approximation. The same hierarchy of reaction rates can be seen in Fig.~\ref{fig:W-rho} for the $\mathcal{W}$ functions. Notice that $\rho_\text{start}$ points (where $\lambda$'s rise up from zero in Fig.~\ref{fig:lambda-rho}) differ from $\Lambda$ onset densities for \nLLL\ and from $\Xi^-$ onset densities for \nXLX, since the conditions $\Theta_{n\Lambda\leftrightarrow \Lambda\Lambda}>0$ and $\Theta_{n\Xi\leftrightarrow \Lambda\Xi}>0$ can be satisfied only for high enough $n_\Lambda$ and $n_\Xi$.

\subsection{OME vs $W$ exchange}
\label{sec:lambdas-OMEvsW}

Let us compare the reaction rates derived using the OME interaction to what one has for the contact $W$ exchange interaction. Only two processes among the {considered ones} go via $W$ exchange, \npLp\ and \LnXp. Here we focus on the former one. For {simplicity} we use the non-relativistic matrix element \cite{LO2002,vDD2004,NO2006} 
\begin{equation}
\label{eq:Mfi-contact}
\left\langle \left| \mathcal{M}_\text{\npLp}^W \right|^2 \right\rangle = 2 G_\text{F}^2 \sin^2 2\theta_\text{C} m_n m_p^2 m_\Lambda \chi_\text{\npLp},
\end{equation}
where $\sin\theta_\text{C} = 0.231$, $\theta_\text{C}$ is the Cabibbo angle, and $\chi_\text{\npLp} = 1 + 3|c_A^{np}|^2 |c_A^{p\Lambda}|^2 \approx 3.47$ with the axial coupling constants $c_A^{np} = -1.26$ and $c_A^{p\Lambda} = -0.72$ \cite{LO2002,vDD2004}.\footnote{
We emphasise that here $\langle|\mathcal{M}_{12\to 34}|^2\rangle$ is the matrix element, squared, summed over the final spin states, and averaged over the initial spines. Our notation should not be confused  with notations used in~\cite{LO2002} and~\cite{vDD2004}.
} We use here the bare baryon masses, {as in} \cite{LO2002,vDD2004,NO2006}. The matrix element in Eq.(\ref{eq:Mfi-contact}) does not depend on angles between the reacting particles momenta, so Eq.~(\ref{eq:IAJ-J}) yields $\mathcal{J} = \langle|\mathcal{M}_\text{\npLp}|^2\rangle$. The reaction rate in {the case} of $W$ exchange can be expressed in the same form as for {the OME} interaction (Eq.~\ref{eq:lambda1234}). Using Eq.~(\ref{eq:J-W}), one finds that $\lambda_\text{\npLp}$ {obtained via} the $W$ exchange is given by Eq.~(\ref{eq:lambda1234}) with 
\begin{multline}
\label{eq:WlpnpContact}
\mathcal{W}_\text{\npLp}^W = 
\\
\frac{\sin^2 2\theta_\text{C}}{8 s_\text{\npLp}} \frac{m_n}{m_N} \frac{m_\Lambda}{m_N} \left( \frac{m_p}{m_N} \right)^2 \chi_\text{\npLp} \approx 0.10.
\end{multline}
This is $7-15$ times less than for \npLp\ using the OME interaction, 
{in accordance} with the results {obtained in} \cite{vDD2004}. 

\begin{figure}
\includegraphics[width=\columnwidth]{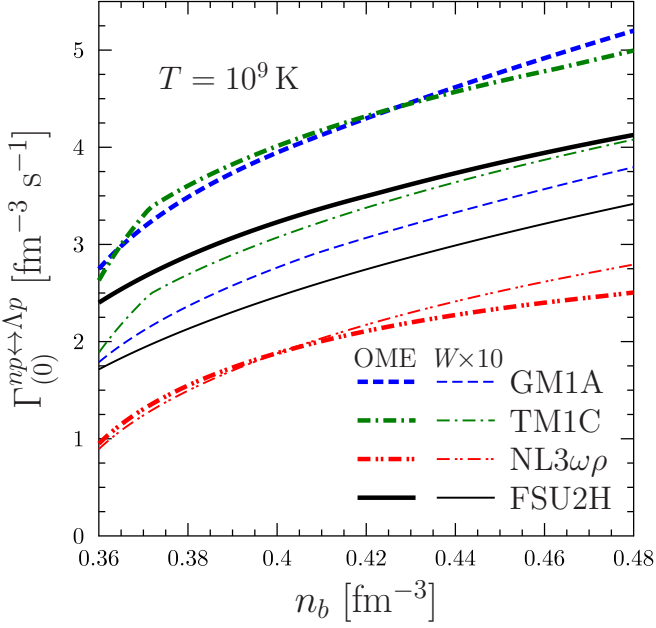}
\caption{\label{fig:OMEvsW} Equilibrium reaction rates $\Gamma_0$ {for} the \npLp\ process. Thick lines are for the OME channel, thin lines are for the contact $W$ exchange channel multiplied by $10$.}
\end{figure}

To compare our {results} with \cite{vDD2004}, we calculate {the equilibrium rate} of reactions for the \npLp\ process, $\Gamma_{(0)}^\text{\npLp}$, which is related to the subthermal reaction rate $\lambda_\text{\npLp}$ according to
\begin{equation}
\label{eq:G0-lambda}
\Gamma_{(0)}^{12\leftrightarrow 34} = \frac{3 k T}{2\pi^2} \lambda_{12\leftrightarrow 34}.
\end{equation}
We plot these rates for each EoS model from Sec.~\ref{sec:EoSs} in Fig.~\ref{fig:OMEvsW}. This figure is similar to figure~7 from \cite{vDD2004}: our thick lines correspond to their solid line ($\Gamma_{(0)}^\text{\npLp}$ using OME), and our thin lines correspond to their dotted line ($10\times \Gamma_{(0)}^\text{\npLp}$ using contact $W$ exchange). 
As expected, the OME interaction yields the equilibrium rate $\sim 10$ times greater than the $W$ exchange. 
But, surprisingly, our {calculations} give $\Gamma_{(0)}$ systematically $\gtrsim 4$ times lower than in \cite{vDD2004}, both for {the OME and the $W$} exchange channels.

\subsection{Comparison of the reaction rates and $\lmax$}
\label{sec:lambdas-Topt}

\begin{figure}
\includegraphics[width=\columnwidth]{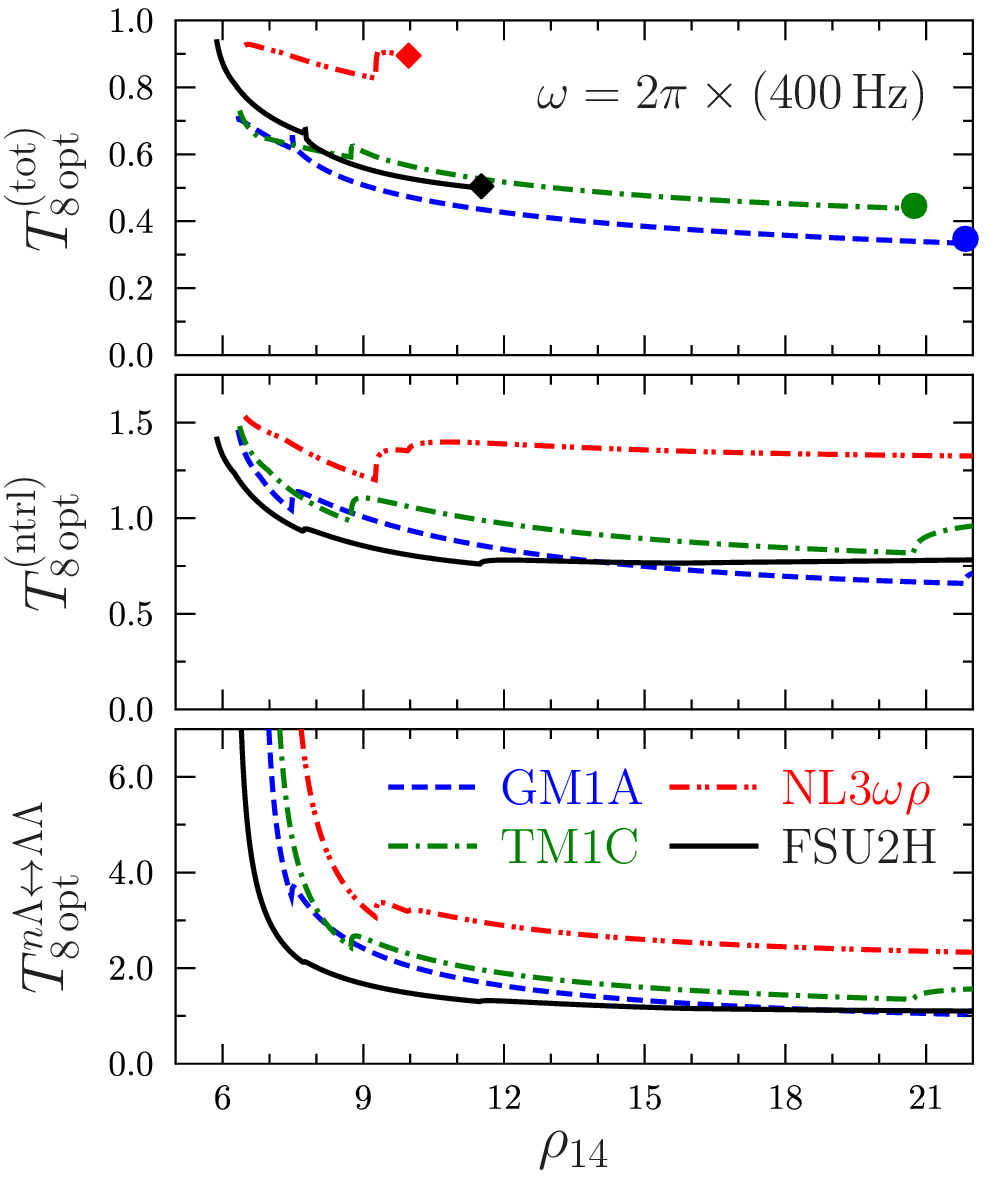}
\caption{\label{fig:Topt} Optimum temperatures for the bulk viscosity at $\omega=2\pi\times (400\,$Hz) {assuming}:
nonsuperfluid and nonsuperconducting matter (\textit{top}; see Eq.~\ref{eq:Topt-tot}); 
strong superfluidity of charged baryons and nonsuperfluid neutral ones
(\textit{middle}; see Eq.~\ref{eq:Topt-ntrl}), and 
optimum temperature for the case when only the reaction 
\nLLL\ operates (\textit{bottom}; see Eq.~\ref{eq:Topt-nLLL}). 
In the top panel diamonds and circles mark the $\Sigma^-$ and $\Xi^0$ onsets, correspondingly, where the set of reactions included in the total $\lambda$ becomes incomplete. 
In each case the curves are plotted at $\rho\geqslant 1.01\rho_\text{start}$ to avoid discontinuities.}
\end{figure}

Now we are able to answer the question from the end of the previous section, namely, how close the total rate $\lambda$ (the sum of all $\lambda_{12\leftrightarrow 34}$, see Eq.~\ref{eq:lambTotDef}) can be to the optimum rate $\lmax$. {To answer it,} we need to calculate ``the optimum temperature'', at which the bulk viscosity reaches its maximum,
\begin{equation}
\label{eq:Topt-tot}
T_\text{opt}^\text{(tot)} = 10^8\,\text{K} \times\sqrt{\omega_4} \sqrt{\frac{\lmax\bigr|_{\omega_4=1}}{\lambda\bigr|_{T_8=1}}},
\end{equation}
and check whether {such a temperature can exist} in the NSs {we are interested in}. The upper panel in Fig.~\ref{fig:Topt} {shows} $T_\text{opt}^\text{(tot)}$ at $\omega=2\pi\times (400 \, {\rm Hz})$
as a {function of  density}. 
The chosen frequency is typical for 
those NSs in LMXBs, 
which could be 
subject to the  r-mode instability 
\cite{Haskell2015}. We plot the curves up to 
the points of $\Sigma^-$ or $\Xi^0$ onset, where the set of considered reactions {becomes} incomplete. A typical optimum temperature value is within the range of $(0.5-1)\times 10^8\,$K, that {might be} close to the typical internal temperature of NSs in LMXBs. Thus application of our hyperon bulk viscosity to the problem of r-mode stability has some chances for success.

{Up to this point  we were considering  only a non-superfluid (non-paired) 
nucleon-hyperon matter}. Baryon pairing is known to suppress reaction rates dramatically \cite{HLY2002} and affects substantially hydrodynamics of NS matter, in particular, the relation between the bulk viscosity(-ies) and the reaction rates \cite{GK2008}. Anyway, here we do not account for the latter effect, and use non-superfluid $\lmax$ to compare {it} with suppressed reaction rates. As is widely accepted \cite{PageReview2013,SedrClark2018}, 
neutral baryons in the NS cores have lower pairing critical temperatures than the charged ones. 
Thus, the first step will be to suppress processes involving $p$, $\Xi^-$, etc. 
A conservative way to do that is to switch off completely 
all the processes involving charged baryons 
(in our case \npLp, \LnXp, and \nXLX). 
Then one can introduce the optimum temperature {for only reactions with neutral particles}
\begin{equation}
\label{eq:Topt-ntrl}
T_\text{opt}^\text{(ntrl)} = 10^8\,\text{K} \times \sqrt{\omega_4} \sqrt{\frac{\lmax\bigr|_{\omega_4=1}}{(\lambda_\text{\nnLn} + \lambda_\text{\nLLL})\bigr|_{T_8=1}}}.
\end{equation}
It is plotted in the middle panel of Fig.~\ref{fig:Topt}. It appears to be about $1.5$ times higher than in the unpaired case, $T_\text{opt}^\text{(ntrl)} \sim (0.8-1.5)\times 10^8\,$K. 
{One can go further and suggest}
that the critical temperature of $\Lambda$'s is significantly lower than the neutron critical temperature \cite{Takatsuka2006} since the $\Lambda\Lambda$ interaction is known to be weak \cite{NagaraEvent2001}. 
A way to partially account for pairing of neutral baryons is to switch off the \nnLn\ process, since it is more sensitive to the neutron superfluidity (since more neutrons are involved in the process), and consider \nLLL\ only. Introducing the optimum temperature for this case,
\begin{equation}
\label{eq:Topt-nLLL}
T_\text{opt}^\text{\text{\nLLL}} = 10^8\,\text{K} \times \sqrt{\omega_4} \sqrt{\frac{\lmax\bigr|_{\omega_4=1}}{ \lambda_\text{\nLLL}\bigr|_{T_8=1}}},
\end{equation}
we get the bottom panel of Fig.~\ref{fig:Topt}. The optimum temperature is significantly higher in this case, especially at densities close to the threshold of the \nLLL~process\footnote{
In all these three cases $T_\text{opt}$ tends to infinity in the vicinity of the corresponding $\rho_\text{start}$, but in the former two cases this divergence is insensible at $\rho \geqslant 1.01\rho_\text{start}$, where the curves in Fig.~\ref{fig:Topt} are plotted.
}. A typical hyperon NS {core} with the central density $\sim 3\rho_0$ should be rather hot, $\sim (2-5)\times 10^8\,$K, to achieve the most effective viscous damping in its interiors.

However, even if the regime $\zeta = \zmax$ is not reached in the NS core, the calculated bulk viscosity can significantly affect the r-mode stability, as it is demonstrated in the next section.

\section{R-mode instability windows}
\label{sec:windows}

Considering the r-mode instability windows, we follow the approach {of} \cite{NO2006}. Namely, we focus on the quadruple $l = m = 2$ r-mode, {which is treated} within the non-superfluid non-relativistic hydrodynamics (cf. Sec.~\ref{sec:zeta-lambda}), but with radial density profiles $\rho(r)$, $n_j(r)$, etc., taken from the numerical solution to the Tolman-Oppenheimer-Volkoff equations \cite{OppVol1939,Tolman1939}. The stability criterion {for the} r-mode is
\begin{equation}
\label{eq:stabCrit}
\frac{1}{\tau_\text{GW}(\nu)} + \frac{1}{\tau_\zeta(\nu,\Tg)} + \frac{1}{\tau_\eta(\Tg)} > 0,
\end{equation}
where $\tau_\text{GW} < 0$ is the driving timescale of the instability due to 
{the gravitational} wave emission (Chandrasekhar-Friedman-Schutz instability \cite{Chandra1970,FriSch1978}), $\tau_\zeta > 0$ is the damping timescale due to the bulk viscosity, and $\tau_\eta > 0$ describes damping due to the shear viscosity. These timescales depend on the rotation frequency $\nu$ and the {redshifted internal} temperature $\Tg$ (assumed to be constant over the NS core). The $\nu(\Tg)$ dependence, {for which} the inequality~(\ref{eq:stabCrit}) {becomes} an equality, {corresponds to} the critical frequency curve in the $\nu-\Tg$ plane.
The region of $\nu$ and $\Tg$, 
where the condition~(\ref{eq:stabCrit}) is violated 
(above the critical $\nu$ curve) 
is the r-mode instability window for a NS. 
{Observing} NSs with frequency and temperature in this domain 
is highly unlikely \cite{Haskell2015}.

{The} necessary formulas for $\tau_\text{GW}$ and $\tau_\zeta$ 
can be found in \cite{NO2006}. 
For the latter timescale we use $\zeta$ obtained in the two previous Sections 
(Eqs.~\ref{eq:zetaFin}, \ref{eq:zlmax}, 
supplemented with Eqs.~\ref{eq:lambda1234}, 
\ref{eq:lambda0+Wdef} for required processes). 
{The} derivation of $\tau_\eta$ is given in \cite{LOM1998}. The main contribution to the shear viscosity $\eta$ comes from leptons, $e$ and $\mu$, 
independently of whether baryons are in the normal or in the superfluid state \cite{SchSht2018}.
Moreover, if protons are superconducting, lepton shear viscosity $\eta$ 
is enhanced \cite{SchSht2018,Sht2018}. 
Since the shear viscous damping is mostly important at low temperatures, 
where protons are paired, 
we have to use the ``superconducting'' expression for $\eta$. Luckily, there is an upper estimate for $\eta$ 
which is independent of pairing properties (the ``London limit'', $T_{cp} \gg 10^9\,$K; see \cite{Sht2018} for details and the analytic expression).

\begin{figure*}
\includegraphics[width=0.9\textwidth]{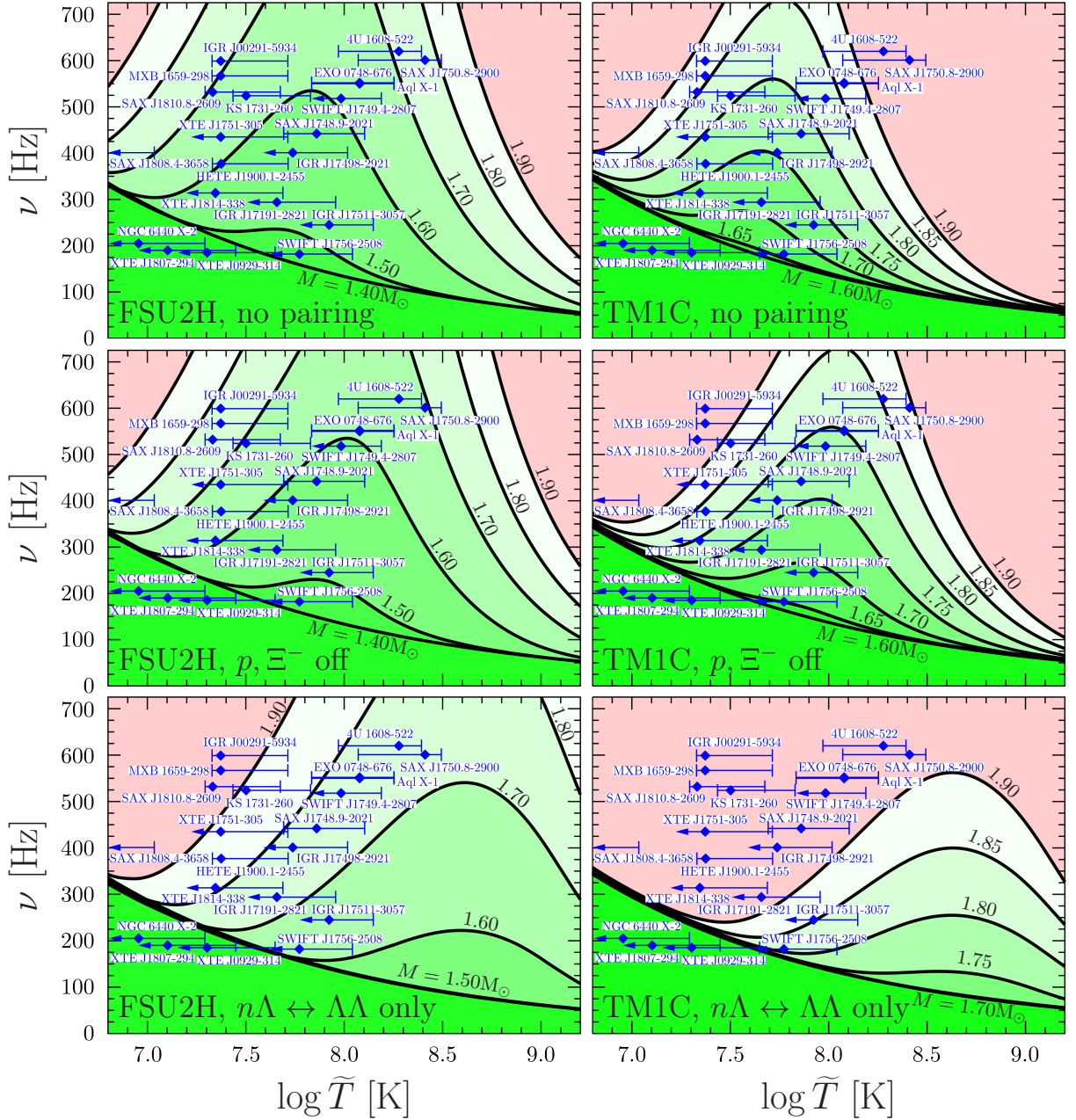}
\caption{\label{fig:windows} {Example of r}-mode critical frequency curves for FSU2H (left) and TM1C (right) EoSs
for different neutron star masses (shown near the curves).
For each mass, the unstable region of the rotation frequency $\nu$ and the redshifted internal temperature $\Tg$ (the instability window) is above the curve. On each plot the lowest curve effectively represents {the} critical curve in the absence of hyperons. The upper plots show the instability windows when all the processes in the set~(\ref{eq:weakReact}) operate (no account for baryon pairing). The middle plots illustrate what happens if one {switches} off all reactions involving charged baryons 
(conservative treatment of $p$ and $\Xi^-$ pairing). 
Finally, the bottom plots are for {models that partially account for} $n$ pairing (\nnLn\ is switched off, while \nLLL\ is not affected). 
The blue data points show {the} observed LMXBs with measured $\nu$ and estimated $\Tg$, see \cite{GCK2014} and footnote~\ref{ftnt:LMXBs} for details.}
\end{figure*}

Fig.~\ref{fig:windows} {shows} the instability windows for various NS models. The top two panels are for the bulk viscosity unaffected by baryon pairing (all five processes in Eq.~\ref{eq:weakReact} {operate}). We restrict ourselves {to NS} with $M\leqslant 1.9\,$\Msun\ to avoid the appearance of $\Sigma^-$ hyperons. Similarly to Sec.~\ref{sec:lambdas-Topt}, we consider the $p$ and $\Xi^-$ pairing effects {excluding} all reactions involving these particles (two middle panels in Fig.~\ref{fig:windows}), and {simulating} $n$ pairing effects by excluding the reaction \nnLn\ (bottom panels in Fig.~\ref{fig:windows}). 
However, in all plots we use the expressions~(\ref{eq:zetaFin}), (\ref{eq:zlmax}) for a relation between the reaction rates and the bulk viscosity, i.e. we ignore influence of pairing effects on hydrodynamics of the core matter (similar to Sec.~\ref{sec:lambdas-Topt}). 
Figure~\ref{fig:windows} presents the instability windows for FSU2H and TM1C EoSs only. {Plots for GM1A EoS are similar to those for FSU2H EoS.}
In turn, \nliiiwr\ critical frequency curves resemble the ones for TM1C, except for the substantially greater $\Lambda$ onset mass (see Table~\ref{tab:astro}) and a slower growth with increasing $M$. 
For instance, \nliiiwr\ NS with $M=2.55\,$\Msun\ and TM1C one with $M=1.9\,$\Msun\ have almost the same stable $\nu,\Tg$-regions. The latter difference is due to the fact that \nliiiwr\ has 
a {smaller} hyperon fraction than the other three EoSs that we use.

Three main conclusions can be made from {inspecting} Fig.~\ref{fig:windows}. 
First (obvious), is that different EoS models yield different instability windows for the same $M$. 
However, the shape of the critical frequency curve is similar in all cases. 

Second, {the} top of the critical curve is reached at a temperature of the order of the corresponding 
optimum temperature $T_\text{opt}$: $\Tg \sim T_\text{opt}$ (see Sec.~\ref{sec:lambdas-Topt}).
Thus, $T_\text{opt}$ appears to be a good estimate of a NS internal temperature at which r-modes 
are {the most stable}.

Finally, the third {conclusion is} that for {all} EoSs considered above 
a high enough mass can close the instability window in {most} of the area shown in the Figure (except for the right bottom plot). This area is important since it contains the observed sources (LMXBs) that are {difficult to reconcile with current models of r-mode oscillations of NSs} (see e.g. \cite{Haskell2015,GCK2014}). They are shown in Fig.~\ref{fig:windows} by blue data points.\footnote{
\label{ftnt:LMXBs} These sources are the same as in \cite{GCK2014} but with SAX~J1810.8--2609 added ($\nu$ from \cite{Allen+2018}, $\Tg$ derived using \cite{Bilous+2018}). For all the sources $\Tg$ was derived from the effective surface temperature, inferred from observations, assuming $M=1.4\,$\Msun\ and $R=10\,$km. See \cite{GCK2014} for details.
} All these sources appear to be inside the stability regions for high enough NS masses even if $p$ and $\Xi^-$ are ``frozen'' due to the superfluid gaps. In particular, {for the FSU2H EoS} almost all data points {lie} within the contour defined by NSs with a mass of $1.7\,$\Msun and below  with strongly paired charged particles. This is in contrast to Ref.\ \cite{NO2006}, approach to the weak non-leptonic reactions of which {requires} at least partially non-suppressed processes with charged particles. At variance with Ref.~\cite{NO2006} we however account for the \nnLn\ process, not considered by \cite{NO2006}, which appears to be the main contributor to the bulk viscosity in the case of ``frozen'' charged particles. Another difference {with respect to} \cite{NO2006} is that in {that} paper the maximum of the stability curves occurs at $T \gtrsim 10^9\,$K, while we have the maximum of the critical frequency at $T \sim 10^8\,$K (except, maybe, in the case when only \nLLL\ is operating). This is {a consequence of the fact that} we use the OME interaction to calculate the reaction rates, while \cite{NO2006} used the contact one. 

Of course, leaving \nLLL\ {as} the only operating process is not a good way to study effects of $n$ pairing. When the neutron {superfluidity gap rises}, both \nnLn\ and \nLLL\ reaction rates decrease dramatically (the latter one does {it more slowly} than the former one), and none of them is affected in the regions of the NS core where neutrons are not paired yet. A careful consideration of this phenomenon is beyond the scope of the present paper. 

\begin{figure}
\includegraphics[width=\columnwidth]{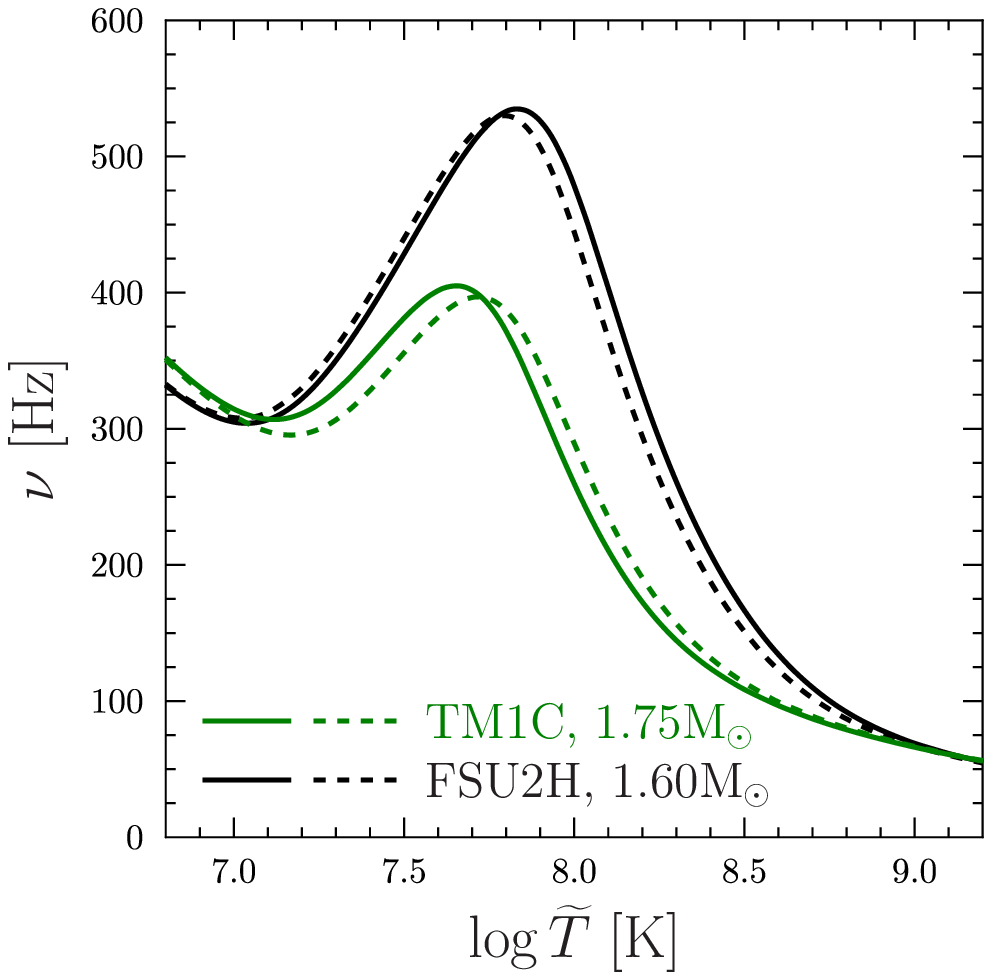}
\caption{\label{fig:windowCheck} Comparison of the critical frequency curves calculated using the exact bulk viscosity (solid lines) and fitting Eqs.~(\ref{eq:zlmaxFit}), (\ref{eq:Wapprox}), (\ref{eq:lambdaApprox}) (dashed lines). The hyperon onset density $\rho_\Lambda$ is adjusted for each EoS. Using Eq.~(\ref{eq:zlmaxFit}), $\zeta_\text{max}$ is multiplied by $1.4$ for FSU2H and by $0.8$ for TM1C. All processes in the set~(\ref{eq:weakReact}) are switched on.}
\end{figure}

In Secs.~\ref{sec:zeta-lambda} and \ref{sec:lambdas} we {provided} the simple approximate expressions for the bulk viscosity. One should {substitute} $\zeta_\text{max}$ and $\lambda_\text{max}$ from Eq.~(\ref{eq:zlmaxFit}) and the reaction rates from combining Eqs.~(\ref{eq:lambda1234}), (\ref{eq:Wapprox}), and (\ref{eq:lambdaApprox}), into Eq.~(\ref{eq:zetaFin}) for the bulk viscosity. The resulting approximation depends on $T$, $\rho$, 
$\rho_\Lambda$ (the density of the hyperons onset), 
and various $\rho_\text{start}$ --- the densities of the reaction thresholds 
(for \npLp\ and \nnLn, $\rho_\text{start}\approx \rho_\Lambda$). 

The value of $\rho_\Lambda$ is fixed for a given EoS but $\rho_\text{start}$ should be accurately adjusted 
for each EoS model in order to obtain a fit that reproduces the instability windows for this EoS. 
Strictly speaking, the parameter $\zeta_0$ in the fitting expression~(\ref{eq:zlmaxFit}) for the maximum bulk viscosity is also very important. While we provided the value $\zeta_0 = 6.5\times 10^{30}\,\text{g}\,\text{cm}^{-1}\,\text{s}^{-1}$ averaged over the four EoSs we use here, its actual value should be adjusted for a given EoS. For instance, FSU2H requires $\zeta_0 \approx 1.4\times$the averaged value, and for GM1A, TM1C, and \nliiiwr\ one needs, respectively, correcting factors $1.45$, $0.8$, and $0.55$. With these comments {taken into account}, the described fit of the bulk viscosity reproduces the critical frequency curves from Fig.~\ref{fig:windows} rather {accurately}, as shown in Fig.~\ref{fig:windowCheck}. Higher accuracy can be achieved if one also adjusts the parameter $s$ in Eq.~(\ref{eq:zlmaxFit}).

\section{Conclusion}
\label{sec:conc}

Let us summarize {the scope of} the present article. First, we calculated the bulk viscosity $\zeta$ for a set {of hyperonic} EoSs. We considered models for which the core is composed of  $npe\mu\Lambda\Xi^-$ matter, in contrast to most of the previous works \cite{LO2002,HLY2002,vDD2004,NO2006} (see, however \cite{ChatBand2006}). We consider the full set of {weak} non-leptonic processes (Eq.~\ref{eq:weakReact}), operating in such NS cores and {generating} $\zeta$. Three of them, \nLLL, \nXLX, and \LnXp, are considered for the first time. The rates $\lambda_{12\leftrightarrow 34}$ for these processes are calculated using the relativistic OME interaction, {as in} Ref.\ \cite{vDD2004} (see Eqs.~(\ref{eq:lambda1234}), (\ref{eq:lambda0+Wdef}), and Appendices~\ref{app:Mfi}, \ref{app:Iang}). {Expressions for} $\zeta$ and $\lambda$'s {are} derived within the non-superfluid hydrodynamics (Eqs.~\ref{eq:zetaFin} and \ref{eq:zlmax}, which {are} appropriate for an arbitrary hyperon composition). 

Second, we calculated the r-mode instability windows following the approach of \cite{NO2006}. We show that {the} positions of the critical frequency curve 
{maxima} are shifted to {lower} temperatures compared to {previous} calculations (cf. Fig.~\ref{fig:windows} and, e.g., Ref.\ \cite{NO2006}), even if we assume strong pairing of charged baryons and moderate pairing of neutral particles in the core. This {is due to the fact that} we calculated the reaction rates using OME interaction instead of the contact $W$ exchange, as Ref.\ \cite{NO2006} did. 

Third, we derived simple approximations for $\zeta$ and $\lambda$'s as a function of $\rho$. {Namely,} for each $\lambda_{12\leftrightarrow 34}$ one may use Eqs.~(\ref{eq:lambda1234}), (\ref{eq:Wapprox}), (\ref{eq:lambdaApprox}) together with the parameters from Tables~\ref{tab:Wapprox}, \ref{tab:lambdaApprox} [or Eqs.~(\ref{eq:lambda1234}), (\ref{eq:lambda0}), and (\ref{eq:Wapprox}) if one wants to specify all particle fractions]. {In turn, to calculate} $\zeta$ one may use Eqs.~(\ref{eq:zetaFin}) and (\ref{eq:zlmaxFit}) together with the approximations for $\lambda$'s. However, {this approximation should be used with caution: if} one wants to reproduce {the} r-mode critical curve for some specific hyperonic EoS, {one} has to adjust the parameters $\zeta_0$ and $\rho_\Lambda$ to this EoS accurately; see the end of Sec.~\ref{sec:windows} and the caption to Fig.~\ref{fig:windowCheck} for an illustration. The value of $\zeta_0$ given in Sec.~\ref{sec:zeta-lambda} is just a rough averaging, appropriate for phenomenological NS models without the detailed hyperon microphysics.

{We would like to point out four limitations of the work presented here}: (i) simplified calculation of the reaction rates; (ii) restricted hyperonic composition; (iii) almost no account for baryon pairing; (iv) simplified calculation of r-mode instability windows.

(i) The first deficiency {in the} $\lambda_{12\leftrightarrow 34}$ calculation is {that} we consider only the lightest meson exchange. In our cases the lightest meson is $\pi$ ($139\,$MeV) for \npLp, \nnLn, \nXLX, and \LnXp, and $K$ ($494\,$MeV) for \nLLL. Both {of} them are pseudoscalar mesons responsible for the long-range interaction. On the one hand, the long-range interaction is typically the most important in rough, first-order approximations, and the up-to-date NS physics does not 
{necessitate} {very} precise calculations of $\lambda$'s. On the other hand, typical distance between the baryons in the NS core is $\lesssim 1\,$fm, while at such distances the transition potential for weak non-leptonic processes strongly deviates from the OME model (at least in atomic hypernuclei \cite{IM2010,PerOb+2013}). So, it is unclear whether the OME interaction model is sufficient for the astrophysical purposes or not. 

Typically, {accounting} for the heavier mesons (first of all, $\rho$ with the mass $770\,$MeV) yields an effect of a factor of few. For decay rates of the hypernuclei, the rates calculated using the $\pi$ exchange only (disregarding the short-range correlations, form factors and final state interactions) are 2--3 times lower than what {is} obtained using many {meson approach} \cite{PRB1997,PRB2002}. In the context of NSs, a comparison of $\pi$ and $\pi+\rho$ exchanges was performed by {Friman and Maxwell} \cite{FM1979} for the neutrino pair bremsstrahlung from $nn$ scattering, $n + n \to n + n + \nu + \tilde{\nu}$. Their result is {that} $\pi$ exchange yields the rate 2--5 times greater than in case of $\pi+\rho$ exchange. A similar effect was obtained using the realistic $T$-matrix instead of one $\pi$ exchange (see the review \cite{SchSht2018} for details).

Another deficiency is our simplistic treatment of the in-medium effects on the meson propagator $D_M$, mainly the pion one ($M = \pi$). As described in Sec.~\ref{sec:lambdas-gator}, the expression~(\ref{eq:DMforUse}) we adopt for the propagators {allows} us to account for the s-wave part of the polarization operator $\Pi$ (in a rather simplistic way), but it provides no account for the p-wave part of $\Pi$. This means that we {underestimate} $D_M$, and, consequently, {also} $\lambda$'s. Different {calculations} of the in-medium modified propagators are divergent \cite{SchSht2018}, the most impressive result is {that} it can increase the reaction rate up to several orders of magnitude \cite{Migdal+1990,Vos2001}.

All in all, are our reaction rates {under or over-estimated}? If the in-medium effects on $D_M$ are close to results {obtained in} \cite{Vos2001}, our $\lambda$'s are surely {underestimated}. If the in-medium effects are not so dramatic, {the situation is unclear}. However, it seems more likely that the effects of $D_M$ in-medium renormalization are stronger than the influence of heavy mesons, so one can expect {that} the reaction rates are higher than {the ones we obtain}.

(ii) Throughout our work we have focused on a $\Lambda\Xi^-$ hyperon composition. For a {number} of EoS models, $\Sigma^-$ {appears} in the core (for instance, in deep layers of massive \nliiiwr\ and FSU2H stars; see also \cite{Prov+2018,TolosCool2018,Fortin+2017}). The relation between $\zeta$ and $\lambda$ inferred in Sec.~\ref{sec:zeta-lambda} is still true in this case, but the total rate $\lambda$ should include the rates of weak non-leptonic processes involving $\Sigma^-$, and may deviate from the $\Lambda\Xi^-$ case. The expressions for the rate $\lambda_{12\leftrightarrow 34}$, given in Sec.~\ref{sec:lambdas}, are applicable for an arbitrary weak non-{leptonic} process $12 \leftrightarrow 34$ operating via the pseudoscalar meson exchange. {However, finding the necessary coupling constants in the literature is not an easy task.}

(iii) The main {limitation} of our work is {that we do not account for baryon pairing}. First of all, it affects the reaction rates. It can be accounted for by introducing {reduction} factors $\mathcal{R}$ \cite{HLY2002}. Some of them are already calculated and analytically approximated, some of them (in particular, $\mathcal R$ for \nnLn\ in the case of $n$ pairing) are {available, but} still not published. We emphasize that a rough account for $\mathcal{R}$'s {via excluding processes involving paired baryons is too simplistic and may be misleading}. Second, baryon superfluidity affects the relation {between} the bulk viscosity and the reaction rates. Moreover, {the number} of kinetic coefficients named ``the bulk viscosity'' increases. These effects were studied in detail by \cite{GK2008,KG2009}. Third, superfluidity affects the r-mode hydrodynamics. Several attempts to explore this effect were {made} \cite{LeeYosh2003,HaskAnd2010,KG2017,DGK2019}, but it is currently an unsolved problem.

(iv) The previous paragraph partially overlaps with the {last limitation} we would like to address, {that is} the simplistic calculation of the r-mode critical frequency curves. Besides the fact that the damping and driving timescales (see Eq.~\ref{eq:stabCrit}) differ in the presence of pairing, the ``$\tau$-approach'' to the critical $\nu$ curve itself is just an estimate. It is widely accepted as it is rather accurate in the non-paired case, but in the presence of pairing this approach should be revisited \cite{DGK2019}. Next, we calculate the damping timescale $\tau_\zeta$ due to the bulk viscosity employing the same approach as in Ref.~\cite{NO2006}. In particular, we used their fitting formula for the angle averaged $(\diver\vec{u})^2$, which was fitted to {NS models obtained using their specific collection of EoSs}. It can be less accurate for our {choice of} EoSs. {Finally}, we use {non-relativistic hydrodynamics}, which is also inaccurate in NSs. 

{Improving the model presented in this work and overcoming, in particular, the limitations} (ii) and (iii), i.e. including more hyperon species and calculating 
the $\mathcal{R}$-factors that are currently unavailable, {will be the} subject of our future work.

\begin{acknowledgments}
%
This work is supported in part by the Foundation for the Advancement of Theoretical Physics and mathematics ``BASIS'' [Grant No. 17-12-204-1 (M.E.G.) and 17-15-509-1 (D.D.O.)] and by RFBR Grant No. 18-32-20170 (M.E.G.). D.D.O. is grateful to N.~Copernicus Astronomical Center for hospitality and perfect working conditions. This work was supported in part by the National Science Centre, Poland, grant 2018/29/B/ST9/02013 (P.H.), and grant 2017/26/D/ST9/00591 (M.F.). We thank E.E.~Kolomeitsev and P.S.~Shternin for valuable discussions. 
\end{acknowledgments}

\appendix

\section{Coefficients in Eqs.~(\ref{eq:XYZ})}
\label{app:Mfi}

Let us introduce dimensionless variables
\begin{equation}
\label{eq:AlphaBetaX}
\alpha_j = \frac{\mLand{j}}{m_M}, 
\quad 
\beta_j = \frac{\mDir{j}}{m_M}, 
\quad 
x_j = \frac{\pF{j}}{m_M}.
\end{equation}
In these notations the coefficients in Eq.~(\ref{eq:XYZ-X}) take the form:
\begin{multline}
\label{eq:X0}
X_0 = g_{24} \left( 
  2\alpha_2\alpha_4 - 2\beta_{2}\beta_{4} - x_{2}^2 - x_{4}^2 
\right) 
\\
\times \left[ 
  \left( 
    A_{13}^2 + B_{13}^2 
  \right)
  \left(
    2\alpha_{1}\alpha_{3} - x_{1}^2 - x_{3}^2
  \right) 
\right.
\\
\left.
  + 2\beta_{1}\beta_{3} \left( 
    A_{13}^2 - B_{13}^2
  \right)
\right],
\end{multline}
\begin{multline}
\label{eq:X1}
X_1 = g_{24}^2 \left( A_{13}^2 + B_{13}^2 \right)\left( 
  2 \alpha_{1} \alpha_{3} + 2 \alpha_{2} \alpha_{4} 
\right.
\\
\left.
- x_{1}^2 - x_{2}^2 - x_{3}^2 - x_{4}^2 - 2 \beta_{2} \beta_{4}
\right) 
\\
+ 2 g_{24}^2 \left( A_{13}^2 - B_{13}^2 \right)\beta_{1}\beta_{3},
\end{multline}
\begin{equation}
\label{eq:X2}
X_2 = g_{24}^2 \left( A_{13}^2 + B_{13}^2 \right).
\end{equation}
In Eq.~(\ref{eq:XYZ-Y}) we have:
\begin{multline}
\label{eq:Y0}
Y_0 = g_{14}g_{24}\left( A_{13}A_{23} - B_{13}B_{23} \right) \left[
  \beta_2\beta_3 x_1^2 
\right.
\\
\left.
  + \beta_1\beta_3 x_2^2 
  + \beta_1\beta_3 x_4^2 
  + \beta_2\beta_3 x_4^2 
  - \beta_3\beta_4 x_3^2 
\right.
\\
\left.
  - \beta_3\beta_4 x_4^2 
  + 2\alpha_1\alpha_2\beta_3\beta_4 
  - 2\alpha_1\alpha_4\beta_2\beta_3 
\right.
\\
\left.
  - 2\alpha_2\alpha_4\beta_1\beta_3 
  + 2\beta_1\beta_2\beta_3\beta_4 
\right] 
\\
+ g_{14}g_{24}\left( A_{13}A_{23} + B_{13}B_{23} \right) \left[
  - x_2^2 x_1^2 - x_3^2 x_4^2
\right.
\\
\left.
  - \alpha_1\alpha_2 x_1^2 
  + \alpha_2\alpha_3 x_1^2 
  + \alpha_2\alpha_4 x_1^2 
  - \alpha_1\alpha_2 x_2^2 
\right.
\\
\left.
  + \alpha_1\alpha_3 x_2^2 
  + \alpha_1\alpha_3 x_4^2 
  + \alpha_2\alpha_3 x_4^2 
  + \alpha_1\alpha_4 x_2^2 
\right.
\\
\left.
  + \alpha_1\alpha_4 x_3^2 
  + \alpha_2\alpha_4 x_3^2 
  - \alpha_3\alpha_4 x_3^2 
  - \alpha_3\alpha_4 x_4^2 
\right.
\\
\left.
  + \beta_1\beta_2 x_1^2 
  - \beta_2\beta_4 x_1^2 
  + \beta_1\beta_2 x_2^2 
  - \beta_1\beta_4 x_2^2 
\right.
\\
\left.
  - \beta_1\beta_4 x_3^2 
  - \beta_2\beta_4 x_3^2 
  + 2\alpha_1\alpha_3\beta_2\beta_4 
\right.
\\
\left.
  + 2\alpha_2\alpha_3\beta_1\beta_4 
  - 2\alpha_3\alpha_4\beta_1\beta_2 
  - 2\alpha_1\alpha_2\alpha_3\alpha_4 
\right],
\end{multline}
\begin{multline}
\label{eq:Y1}
Y_1 = g_{14}g_{24} A_{13}A_{23} \left[ 
  \left( \alpha_2-\alpha_3 \right) \left( \alpha_1-\alpha_4 \right)
\right.
\\
\left.
  + \left( \beta_2+\beta_3 \right) \left( \beta_4-\beta_1\right)
\right]
\\
+ g_{14}g_{24} B_{13}B_{23} \left[
  \left( \alpha_2-\alpha_3 \right) \left( \alpha_1-\alpha_4 \right)
\right.
\\
\left.
  - \left( \beta_2-\beta_3 \right) \left( \beta_1-\beta_4 \right)
\right],
\end{multline}
\begin{equation}
\label{eq:Y1p}
Y_2 = Y_1\Bigl|_{1\leftrightarrow 2},
\end{equation}
\begin{equation}
\label{eq:Y2}
Y_3 = g_{14}g_{24}\left( A_{13}A_{23} + B_{13}B_{23} \right).
\end{equation}
In Eq.~(\ref{eq:XYZ-Z}):
\begin{equation}
\label{eq:Z012}
X'_{0,1,2} = X_{0,1,2}\Bigl|_{1\leftrightarrow 2}.
\end{equation}

\section{Transforming Eq.~(\ref{eq:IAJ-J})}
\label{app:Iang}

It is convenient to introduce the dimensionless variables
\begin{equation}
\label{eq:xj}
\vec{x}_j = \frac{\vec{p}_j}{m_M}, 
\quad 
\vec{x} = \frac{\vec{q}}{m_M} = \vec{x}_3 - \vec{x}_1, 
\quad 
\vec{x}' = \frac{\vec{q}'}{m_M} = \vec{x}_3 - \vec{x}_2.
\end{equation}
Similarly, we introduce $x_\text{min,max} = q_\text{min,max}/m_M$ that can be expressed in terms of $x_j = |\vec{x}_j| = \pF{j}/m_M$. The non-weighted angular integral Eq.~(\ref{eq:IAJ-A}) can be written in a dimensionless form $\mathcal{A} = A/m_M^3$ with
\begin{equation}
\label{eq:Adimless}
A = \frac{2(2\pi)^3}{\prod_j x_j} \left( x_\text{max} - x_\text{min} \right) \Theta\left( x_\text{max} - x_\text{min} \right).
\end{equation}

Substituting $\langle\left|\mathcal{M}_{12\to 34}\right|^2\rangle$ from Eq.~(\ref{eq:MfiSqAvSum}) and $D_M$ from Eq.~(\ref{eq:DMforUse}) into Eq.~(\ref{eq:IAJ-J}), we find 
\begin{multline}
\label{eq:Jfin}
\mathcal{J} = G_\text{F}^2 m_\pi^4 \left( X_0 J_0 + X_1 J_1 + X_2 J_2 \right.
\\
\left. + X'_0 J'_0 + X'_1 J'_1 + X'_2 J'_2 \right.
\\
\left. + Y_0 J_3 + Y_1 J_4 + Y_2 J'_4 + Y_3 J_5 \right).
\end{multline}
The dimensionless functions $J_k(x_1, x_2, x_3, x_4)$, $k = 1...5$, are the following:
\begin{multline}
\label{eq:J012}
J_k = \frac{1}{A} \int \prod_j \diff\Omega_j \frac{x^{2k}}{(x^2+1)^2}\delta\left( \vec{x}_1 + \vec{x}_2 - \vec{x}_3 - \vec{x}_4 \right) 
\\
= \frac{\Theta(x_\text{max}-x_\text{min})}{x_\text{max}-x_\text{min}} \int_{x_\text{min}}^{x_\text{max}} \diff x \frac{x^{2k}}{(x^2+1)^2}
\end{multline}
for $k = 0,1,2$, 
\begin{multline} 
\label{eq:J3}
J_3 = \frac{1}{A} \int \prod_j \diff\Omega_j \frac{\delta\left( \vec{x}_1 + \vec{x}_2 - \vec{x}_3 - \vec{x}_4 \right)}{(x^2+1)(x'^2+1)}
\\
= \frac{\Theta(x_\text{max}-x_\text{min})}{x_\text{max}-x_\text{min}} \int_{x_\text{min}}^{x_\text{max}} \frac{\diff x}{(x^2+1)\sqrt{t_1^2(x) - t_2^2(x)}}
\end{multline}
\begin{multline}
\label{eq:J4}
J_4 = \frac{1}{A} \int \prod_j \diff\Omega_j \frac{x^2 \delta\left( \vec{x}_1 + \vec{x}_2 - \vec{x}_3 - \vec{x}_4 \right)}{(x^2+1)(x'^2+1)}
\\
= \frac{\Theta(x_\text{max}-x_\text{min})}{x_\text{max}-x_\text{min}} \int_{x_\text{min}}^{x_\text{max}} \frac{x^2 \diff x}{(x^2+1)\sqrt{t_1^2(x) - t_2^2(x)}},
\end{multline}
\begin{multline}
J_5 = \frac{1}{A} \int \prod_j \diff\Omega_j \frac{x^2 x'^2 \delta\left( \vec{x}_1 + \vec{x}_2 - \vec{x}_3 - \vec{x}_4 \right)}{(x^2+1)(x'^2+1)}
\\
= J_3 + \frac{\Theta(x_\text{max}-x_\text{min})}{x_\text{max}-x_\text{min}} \int_{x_\text{min}}^{x_\text{max}} \diff x \left( \frac{x^2}{x^2+1} \right.
\\
\label{eq:J5}
\left. - \frac{1}{\sqrt{t_1^2(x) - t_2^2(x)}} \right),
\end{multline}
where we use notation of Ref.\ \cite{Maxwell1987}:
\begin{align}
\label{eq:t1}
t_1 &= x_1^2 + x_2^2 - x^2 +1 - 2 x_3 x_4 \cos\theta_1 \cos\theta_2,
\\ 
\label{eq:t2}
t_2 &= 2 x_3 x_4 \sin\theta_1 \sin\theta_2,
\end{align}
with
\begin{align}
\label{eq:cos1}
\cos\theta_1 &= \frac{x_3^2 - x_1^2 + x^2}{2 x_3 x}, 
\\ 
\cos\theta_2 &= \frac{x_2^2 - x_4^2 - x^2}{2 x_4 x}.
\end{align}
For the `exchange' integrals we have
\begin{equation}
\label{eq:Jx5prime}
J'_k = J_k\bigr|_{x_1 \leftrightarrow x_2}
\end{equation}
for $k = 0,1,2,4$, 
that corresponds to $\vec{x} \to \vec{x}'$ 
within the integrals ($J'_k = J_k$ for $k = 3,5$). 
Substituting Eq.~(\ref{eq:Jfin}) into Eq.~(\ref{eq:DeltaGamma-IAJ}), we immediately obtain Eq.~(\ref{eq:lambda1234}).

Reduction of multidimensional integrals to their one-dimensional forms is performed according to the standard technique, see, e.g., Refs.\ \cite{ShapTeuk1983,FM1979,Maxwell1987}. The identities
\begin{equation}
\label{eq:helpForJ}
1 = \int \diff^3\vec{x} \delta\left( \vec{x} + \vec{x}_3 - \vec{x}_1 \right), 
\quad 
\vec{x}' = \vec{x} + \vec{x}_1 - \vec{x}_2
\end{equation}
are helpful \cite{Maxwell1987}. The one-dimensional integrals in the right-hand sides of Eqs.~(\ref{eq:J012}) --- (\ref{eq:J5}) could be simply evaluated, 
both numerically and analytically. 
One can find analytic results in Refs.\ \cite{FM1979,Maxwell1987}.



\end{document}